\documentclass[11pt]{article}

\usepackage{amsmath}
\usepackage{amsfonts}
\usepackage{amssymb}
\usepackage{graphicx}
\usepackage{supertabular}
\usepackage{setspace}
\usepackage{xcolor}
\usepackage{array}
\usepackage{amsthm}
\usepackage{bm}
\usepackage{latexsym}
\usepackage{mathptmx}
\usepackage{hyperref}
\hypersetup{colorlinks=false}

\RequirePackage{cite}%
\renewcommand{\citeleft}{\bgroup\normalfont[}%
\renewcommand{\citeright}{]\egroup}%

\newcommand{\nin}{\noindent}
\newcommand{\nn}{\nonumber}
\newcommand{\be}{\begin{equation}}
\newcommand{\ee}{\end{equation}}
\newcommand{\ba}{\begin{eqnarray}}
\newcommand{\ea}{\end{eqnarray}}
\newcommand{\bal}{\begin{align}}
\newcommand{\eal}{\end{align}}

\newcommand{\e}{{\rm e}}
\newcommand{\dd}{{\rm d}}
\newcommand{\ii}{{\rm i}}
\newcommand{\bb}{\bibitem}

\newcommand{\om}{\omega}
\newcommand{\al}{\alpha}
\newcommand{\la}{\lambda}
\newcommand{\bt}{\beta}

\newcommand{\ga}{\gamma}
\newcommand{\ro}{\rho}

\newcommand{\ta}{\theta}
\newcommand{\Ta}{\Theta}

\newcommand{\De}{\Delta}

\newcommand{\de}{\delta}

\newcommand{\bw}{\begin{widetext}}
\newcommand{\ew}{\end{widetext}}
\newcommand{\eps}{\epsilon}

\def\s{\sqrt2}

\def\ri{\rho_{\infty}}
\def\W{Weierstrass elliptic functions}
\def\w{Weierstrass elliptic functions }

\def\aw{Weierstrass elliptic function }
\def\ws{Weierstrass }
\def\abh{black hole }
\def\bh{black holes }
\def\aBH{black hole}
\def\BH{black holes}
\def\S{Schwarzschild}
\def\s{Schwarzschild }
\def\RN{Reissner-Nordstr\"om }

\renewcommand{\theequation}{\arabic{section}.\arabic{equation}}

\begin{document}


\title{{\large\textbf{Light paths of normal and phantom Einstein-Maxwell-dilaton \BH}}}

\author{Mustapha Azreg-A\"{\i}nou
\\Ba\c{s}kent University, Department of Mathematics, Ba\u{g}l\i ca Campus, Ankara, Turkey}
\date{}

\maketitle

\begin{abstract}
Null geodesics of normal and phantom Einstein-Maxwell-dilaton black holes are determined analytically by the \W. The \abh parameters other than the mass enter, with the appropriate signs, the formula for the angle of deflection to the second order in the inverse of the impact parameter allowing for the identification of the nature of matter (phantom or normal). Such identification is also possible via the time delay formula and observation of relativistic images. Scattering experiences may favor \bh of Einstein-anti-Maxwell-dilatonic theory for their high relative discrepancy with respect to the \s value. For the cases we restrict ourselves to, phantom black holes are characterized by the absence of many-world and two-world null geodesics.

\vspace{3mm}

\nin {\footnotesize\textbf{PACS numbers:} 04.20.Gz, 04.50.Gh, 04.70.--s, 04.70.Bw, 98.62.Sb, 04.20.Dw}

\vspace{-3mm} \nin \line(1,0){430} 
\end{abstract}

\section{Introduction}

Most experimental settings for testing gravitational theories are designed to evaluate trajectories of light rays. Accuracy in this field is a growing interest. From this point of view, the leading experimental settings are aiming to achieve high accuracies beyond the known first order level and to reach a
sensitivity of 1 part in $10^9$ in measuring the Eddington parameter $\ga$~\cite{bl}, which is an important parameter in post-Newtonian formalism.

On the theoretical front, workers have been striving to evaluate exactly light paths using (hyper)elliptic functions mainly the \w denoted by $\wp$~\cite{Wang,Whittaker}. On the one hand, this has provided answers to some open questions, for instance, whether the cosmological constant could be a cause of the Pioneer anomaly~\cite{Hackmann}, has raised the question of whether lensing could be used as a test of the cosmic censorship (CC)~\cite{CC1} (much work on testing the CC has been done in~\cite{CC2,CC,del2}), and has lead to the discovery new light paths, the Pascal Lima\c{c}on trajectories for \bh with cosmological constant~\cite{Cruz}. On the other hand, the analytical solutions derived so far, Refs.~\cite{Ha}-\cite{Ga} to mention but a few, could be useful for any of the experimental settings aiming to test gravitational theories. Moreover, they provide new academic techniques for tackling the motion of massive and massless particles in the geometries of various gravitational fields, may serve as references for testing the accuracy of numerical methods~\cite{ex} and provide unique benchmarks for testing and improving perturbation and decomposition methods~\cite{Ad}. For that purpose it is very helpful to have relatively simple solutions.

In case of spherical symmetry, one of the equations governing geodesic motion reduces to
\begin{equation}\label{1.1}
    \Big(\frac{\dd r}{\dd \phi}\Big)^2 = P(r)
\end{equation}
where $P(r)$ is a polynomial function of the radial variable $r$, the parameters of the solution and the constants of motion. Depending on the dimension of the space-time, $P(r)$ may be reduced, as described in~\cite{met}, to a polynomial of degree 3 or 5. We are interested in the former case and we assume that~\eqref{1.1} is brought to
\begin{equation}\label{1.2}
    \Big(\frac{\dd \ro}{\dd \Ta}\Big)^2 = 4\ro^3-g_2\ro-g_3
\end{equation}
by coordinate transformations. Here $g_2,g_3$ depend on the parameters of the solution and the constants of motion. So far no special terminology has been introduced to simplify notations and expressions. We introduce the following terminology to describe~\eqref{1.2} and the related polynomial and coordinates. We shall call~\eqref{1.2} \ws differential equation, $w(\ro)= 4\ro^3-g_2\ro-g_3$ \ws polynomial and ($\ro,\Ta$) \ws coordinates.

We bring~\eqref{1.1} to~\eqref{1.2} by  a series of coordinate transformations relating $r$ to \ws radial coordinate $\ro$ where $\ro(r)$ is a nontrivial and nonlinear transformation; however, $\Ta(\phi)$ is a linear transformation and in many cases $\Ta=\phi$, where $\phi$ is the azimuthal angle.

Most workers prefer to use the effective potential approach by which they determine all planar trajectories [absorbed paths (captured photons), scattering paths, trapped or confined paths, (un)stable circular paths, spiral paths approaching the circular paths from above and/or below and some other special closed curves]. The method we shall apply is entirely based on the properties of the \ws differential equation, and of its polynomial. We shall develop and use this method, which has been used in~\cite{used2,GV} (and partially used in~\cite{Cruz,used1}), leading to a systematic approach for all problems governed by~\eqref{1.2}. This will allow us to determine all types of trajectories.

None of the papers mentioned above has ever dealt with light paths of normal Einstein-Maxwell-dilaton (EMD) \BH. One of the purposes of this paper is to address this question; the other one is to extend the analysis to that of light paths of phantom \bh of EMD and to draw a comparison between the systems of trajectories for a given ratio of charge to mass squared ($a^2=q^2/M^2$).

In a phantom gravitating field theory one or more of the matter fields appear in the action with an unusual sign of the kinetic term, so that they are coupled repulsively to gravity. In the case of ``phantom EMD" theory, which is also a short term for the theory, we may have a number of ways the matter fields are coupled to gravity: Einstein-anti-Maxwell-anti-dilaton, Einstein-Maxwell-anti-dilaton, Einstein-Maxwell-dilaton and so on. The presence of phantom fields continues to receive support from both collected observational data~\cite{Koma} and theoretical models~\cite{dyn}.

The static, spherically symmetric black hole solutions to EMD theory with phantom Maxwell and/or dilaton field were derived, and their causal structure was analyzed, among which one finds nine classes of asymptotically flat and two classes of nonasymptotically flat phantom black holes~\cite{phantom}. In a subsequent work~\cite{multi}, these solutions have been generalized to multicenter solutions of phantom EMD. Recently, their thermodynamic properties and stability were investigated too~\cite{thermo}. One of the remaining tasks is to investigate their null geodesics to see how phantom fields may affect the light paths, particularly the angles of deflection, the photon spheres and related lensing effects. Deviations of the angle of deflection from the \s value are generally attributable to extensions in the theory (inclusion of Maxwell fields or scalar ones, cosmological constant and so on), departure from spherical symmetry or motion of the solution itself (mostly rotation). In this paper we examine the case due to the inclusion of (anti)-dilatonic and/or (anti)-Maxwell fields.

In Section~\ref{2} we consider the cosh and sinh \abh solutions of the generalized phantom EMD, which depend on three parameters ($M,q,\ga$). We evaluate, and discuss, the angle of deflection $\de\phi$ to the second order of approximation in the inverse of the impact parameter as a function of the \abh three parameters. Figures, relying on exact formulas, depicting $\de\phi$ for phantom and normal \bh are plotted against the \s angle of deflection for different values of the parameters. The relative discrepancy is discussed and plotted showing high values from some set of the parameters. The time delay is also evaluated.

In Section~\ref{3} we introduce the \w and use and develop the method based on the \ws polynomial to determine exactly all kinds of null geodesics to any spherically symmetric geometry, provided the equation of (planar) motion of light rays may be brought to~\eqref{1.2}. Applications are considered in Section~\ref{3d} and in Sections~\ref{5} and~\ref{4}. In Subsection~\ref{3d} we consider the strong field limit and relativistic images and derive an analytic reference equation for the log-formula for the angle of deflection, which applies to any geometry provided the light motion or a plane projection of it is described by~\eqref{1.2}. In Section~\ref{5} we consider the case $\ga=1$ and show that the problem of determining the null geodesics of normal \RN \bh by the method based on the \ws polynomial, which was initiated in~\cite{GV}, is tractable analytically and extend the analysis to phantom \RN \bh upon applying the results of Section~\ref{3}, and in Section~\ref{4} we consider the case $\ga=0$ and determine all the null geodesics of the phantom cosh and normal sinh EMD \bh by mere comparison to the work done in Section~\ref{3}. In Sections~\ref{3} to~\ref{4}, we do not aim to go into the details of each null geodesic motion; rather, we present a general procedure (Section~\ref{3}) by which we discuss some type of null geodesic motions and the nature of existing divergencies and present exact reference and standard formulas for specific geodesics, the angle of deflection, the time delay, and the log-formula. The paper ends with a conclusion section and an two appendix sections.

\setcounter{equation}{0}
\section{The deflection angle of light paths in the cosh-sinh solutions of EMD}\label{2}

The action for EMD theory with phantom Maxwell and/or dilaton field reads
\begin{equation}\label{2.1}
S=-\int \dd ^{4}x \sqrt{-g}\,\;[\mathcal{R}-2\eta_1 g^{\mu\nu}
\partial_{\mu}\varphi\partial_{\nu}\varphi +\eta_2 \e^{2\lambda\varphi}
F_{\mu\nu}F^{\mu\nu}]\,,
\end{equation}
where $\lambda$ is the
real dilaton-Maxwell coupling constant, and $\eta_1=\pm 1$, $\eta_2=\pm 1$. Normal EMD
corresponds to $\eta_2=\eta_1=+1$, while phantom couplings of the dilaton field $\varphi$
or/and Maxwell field $F = \dd A$ are obtained for $\eta_1=-1$ or/and $\eta_2=-1$.

The metrics of the so-called cosh and sinh solutions, derived in~\cite{phantom}, take the form
\begin{align}
\label{2.2}& \dd s^{2}=f_{+}f_{-}^{\gamma}\dd t^{2}
-f_{+}^{-1}f_{-}^{-\gamma}\dd r^{2}
-r^{2}f_{-}^{1-\gamma}\dd \Omega^{2}\\
& F=-\frac{q}{r^2}\,\dd r\wedge \dd t\, ,\;\;
\e^{-2\lambda\varphi}=f_{-}^{1-\gamma}\,,\;\; f_{\pm}=1- \frac{r_{\pm}}{r}\,,\;\;\ga=\frac{1-\eta_1\la^2}{1+\eta_1\la^2} \nn\\
\label{2.3}&\eta_2(1+\eta_1\la^2)<0 \text{ for cosh}\,,\;\eta_2(1+\eta_1\la^2)>0 \text{ for sinh}\\
\label{2.4}& \ga \in (-\infty,-1)\cup[1,+\infty) \text{ if } \eta_{1}=-1\,,\;\ga \in (-1,+1] \text{ if } \eta_{1}=+1
\end{align}
where we have introduced the parameter $\ga$ following the notation of~\cite{thermo}\footnote{For the sinh solution the case $\eta_2(1+\eta_1\la^2)<0$, which would lead to $r_-<0$, is not possible~\cite{phantom}.}.

These are asymptotically flat spherically symmetric black holes of mass $M$, electric charge $q$ and event horizon $r_{+}>0$ related by~\cite{phantom}
\begin{equation}\label{2.6}
M  =\frac{r_{+}+\gamma r_{-}}{2}\,,\;\;q=\pm \sqrt{\frac{1+\gamma}{2}\eta_{2}r_{+}r_{-}}
\end{equation}
where we have substituted, into the original formula of $q$, $1+\eta_1\la^2=2/(1+\ga)$. Since $q$ is real, $r_-$ and $\eta_2(1+\ga)$ must have the same sign. Using this fact in~\eqref{2.3}, we have $r_-<0$ for the cosh solution and $r_->0$ for the sinh one.

As shown in Subsection 4.1 case 2. (d) of~\cite{phantom}, $r=0$ corresponds to a singularity for the cosh solution where geodesics terminate (a Penrose diagram is given in figure 1 of~\cite{phantom}). Similarly, in Subsection 4.3 case 1. (a) (ii) of~\cite{phantom} it is established that, for generic values of $1+\eta_1\la^2$ as this is the case for $\ga=0$ (to which we restrict ourselves in Section~\ref{4}), $r=r_-$ is a null singularity for the sinh solution (a Penrose diagram is given in Figure 3 of~\cite{phantom}). The curvature scalar of~\eqref{2.2} diverges at these two points for $\ga =0$
\begin{equation*}
    \mathcal{R}=-\frac{r_-(r-r_+)}{2r^3(r-r_-)^2}\,.
\end{equation*}

Expressing ($r_+,r_-$) in terms of $M$ and $a^2=q^2/M^2$, one obtains
\begin{align}
\label{2.8}& r_+=2M\,,\;r_-=\eta_2Ma^2\,,\quad \text{if }\ga = 0\\
\label{2.9}& r_{+}=M+\mathcal{M}\,,\;
    r_{-}=\frac{M-\mathcal{M}}{\ga}=\frac{2\eta_2M^2a^2}{(1+\ga)r_+}\,,\;\mathcal{M}=M\sqrt{1-\frac{2\eta_{2}\ga a^2}{(1+\ga)}}\,,\quad \forall \ga \neq -1\,.
\end{align}
[The limit $\ga\to 0$ in~\eqref{2.9} leads to~\eqref{2.8}].

\paragraph*{\textbf{Angle of deflection.}}

The derivation of the angle of deflection is given in Appendix A by
\begin{equation}\label{2.16}
    \delta\phi = \frac{4M}{r_n}+ \big\{-2M[ r_++r_- (2\gamma -1)]+\frac{\pi }{16} [15r_+^2+6 r_-r_+(4 \gamma -1)+r_-^2(16\gamma ^2-1)]\big\}\frac{1}{r_n^2}+O[1/r_n]^3
\end{equation}
where $u_n=1/r_n$ is the point on the light scattering geodesic nearest the origin where $\frac{\dd u}{\dd \phi}(u_n)=0$. Since the values of $\ga$ depend on $\eta_1$ according to (\ref{2.4}) and the sign of $r_-$ is that of $\eta_2(1+\ga)$ by \eqref{2.6}, the deflection angle depends on the type of EMD under investigation. From (\ref{2.8}, \ref{2.9}) one sees that, for both cases $\ga =0$ and $\ga \neq0$, the limit case $q= 0$ corresponds to $r_-=0$ and $r_+=2M$. Thus in the limit $q\to 0$, $\delta\phi$ approaches the value $\delta\phi(r_-=0,r_+=2M)$, which is the \s angle of deflection $\delta\phi_{\text{S}}$:
\begin{equation}\label{2.16a}
    \lim_{q\to 0}\delta\phi = \delta\phi(r_-=0,r_+=2M)=\delta\phi_{\text{S}}= \frac{4M}{r_n}+\frac{(15\pi-16)M^2}{4}\,\frac{1}{r_n^2}+O[1/r_n]^3\,.
\end{equation}
Using this along with (\ref{2.8}, \ref{2.9}) in~\eqref{2.16} we express $\delta\phi$ in terms of the charges ($M,q$) and $\delta\phi_{\text{S}}$
\begin{align}
\label{2.17}&\delta\phi = \delta\phi_{\text{S}}-\frac{\pi M^2a^2}{16}\big[\eta_2\frac{4(3\pi-8)}{\pi}+a^2\big]\,\frac{1}{r_n^2}+\cdots\,,\quad \text{if }\ga =0\\
\label{2.18}&\delta\phi = \delta\phi_{\text{S}}-\eta_2\big[\frac{(\ga -1)[16\ga -\pi(\ga +1)]M|M-\mathcal{M}|+ \pi\ga (7\ga -1)q^2}{8\ga^2}\big]\,\frac{1}{r_n^2}+\cdots\,,\quad \forall \ga \neq -1\\
\label{2.19}& \quad \;=\frac{4M}{r_n}+\bigg\{-2M(M+\mathcal{M})+\eta _2\frac{4(1-2 \gamma ) a^2M^3}{(1+\gamma )(M+\mathcal{M})}\\\nn
& \quad \;+\frac{\pi }{16} \Big[15(M+\mathcal{M})^2+\frac{4(16\gamma ^2-1)a^4M^4}{(1+\gamma
)^2(M+\mathcal{M})^2}+ \eta _2\frac{12(4\gamma -1)a^2 M^2}{1+\gamma }\Big]\bigg\}\frac{1}{r_n^2}+\cdots\,,\quad \forall \ga \neq -1
\end{align}
where we have made use of $M-\mathcal{M}=\eta_2 |M-\mathcal{M}|$ and~\eqref{2.9}. [The limit $\ga\to 0$ in~\eqref{2.18} or in~\eqref{2.19} leads to~\eqref{2.17}]. Thus to the first order of approximation in $1/r_n$ all normal and phantom \bh deflect light paths in the same way with $\delta\phi=4M/r_n+\cdots$. To the second order of approximation in $1/r_n$, the added contribution to the \s one [second terms in (\ref{2.17}, \ref{2.18})] does not depend on the sign of $q$ but depends on the signs of ($\eta_1,\eta_2$).
\begin{figure}[h]
\centering
  \includegraphics[width=0.49\textwidth]{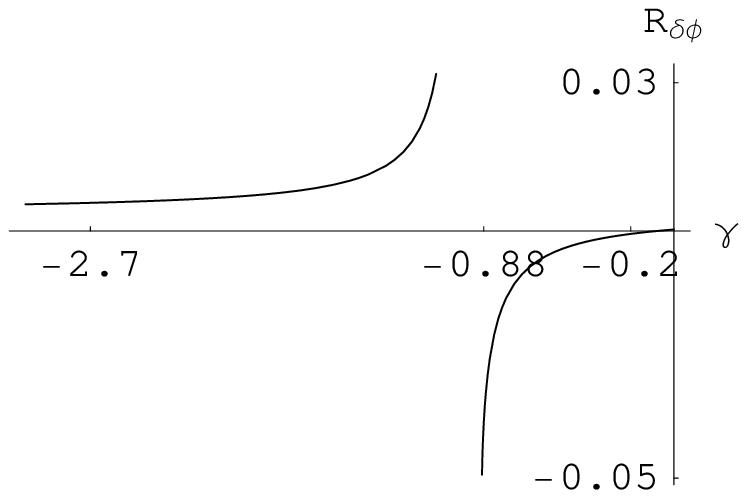}  \includegraphics[width=0.49\textwidth]{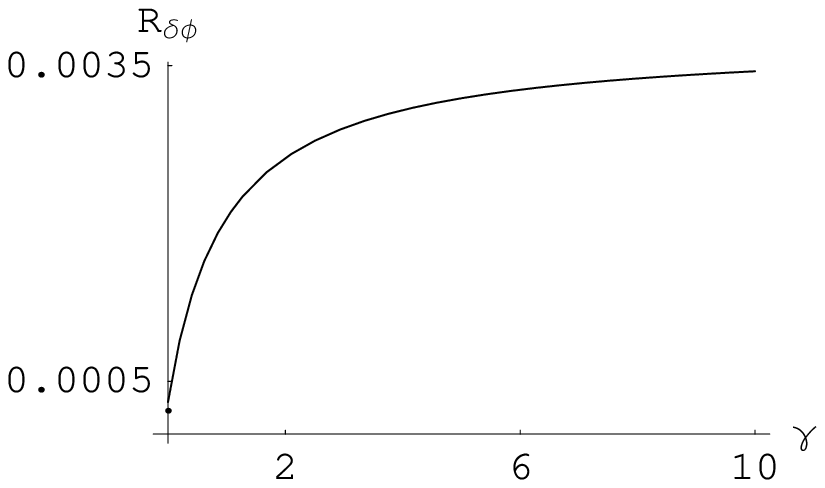}\\
  \caption{\footnotesize{The relative discrepancy $R_{\de\phi}=(\de\phi-\delta\phi_{\text{S}})/\delta\phi_{\text{S}}$, which defines the relative difference of the actual deflection angle with respect to the \s value, is sketched on its domain of definition against $\ga$ for fixed ($M=1,a^2=1/16,u_n=0.05,\eta_2=-1$), $r_n=1/u_n$ is the point on the null geodesic nearest the origin. This is the E-anti-M-(anti)-D case ($\eta_1$ depends on $\ga$). $R_{\de\phi}$ increases on its domain of definition and changes sign for some $\ga_0$ between $-0.1$ and $-0.05$.}}\label{Fig5}
\end{figure}
\begin{figure}[h]
\centering
  \includegraphics[width=0.49\textwidth]{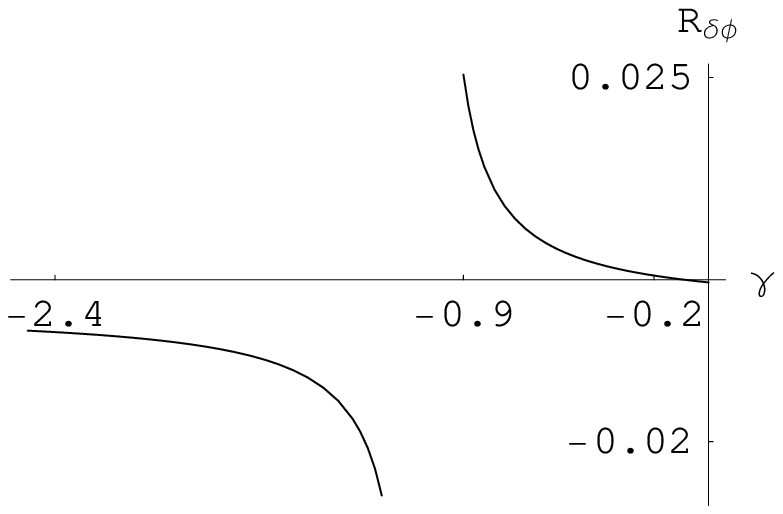}  \includegraphics[width=0.49\textwidth]{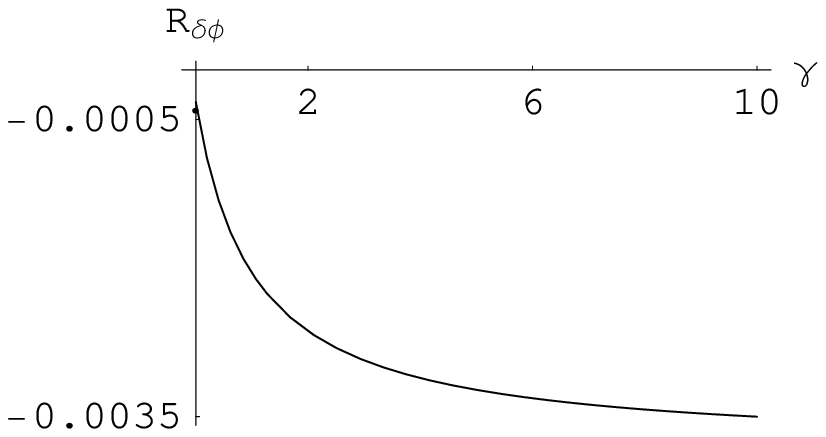}\\
  \caption{\footnotesize{The relative discrepancy $R_{\de\phi}=(\de\phi-\delta\phi_{\text{S}})/\delta\phi_{\text{S}}$, which defines the relative difference of the actual deflection angle with respect to the \s value, is sketched on its domain of definition against $\ga$ for fixed ($M=1,a^2=1/16,u_n=0.05,\eta_2=1$), $r_n=1/u_n$ is the point on the null geodesic nearest the origin. This is the EM-(anti)-D case ($\eta_1$ depends on $\ga$). $R_{\de\phi}$ decreases on its domain of definition and changes sign for some $\ga_0$ between $-0.1$ and $-0.05$.}}\label{Fig6}
\end{figure}
\begin{figure}[h]
\centering
  \includegraphics[width=0.49\textwidth]{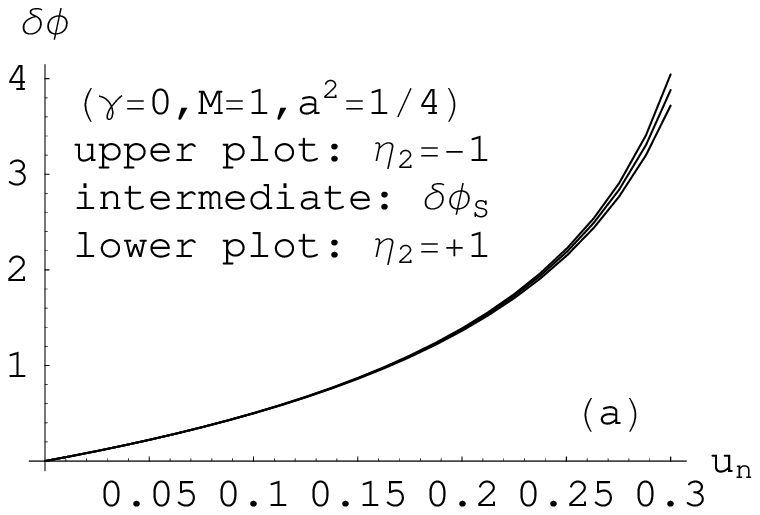}  \includegraphics[width=0.49\textwidth]{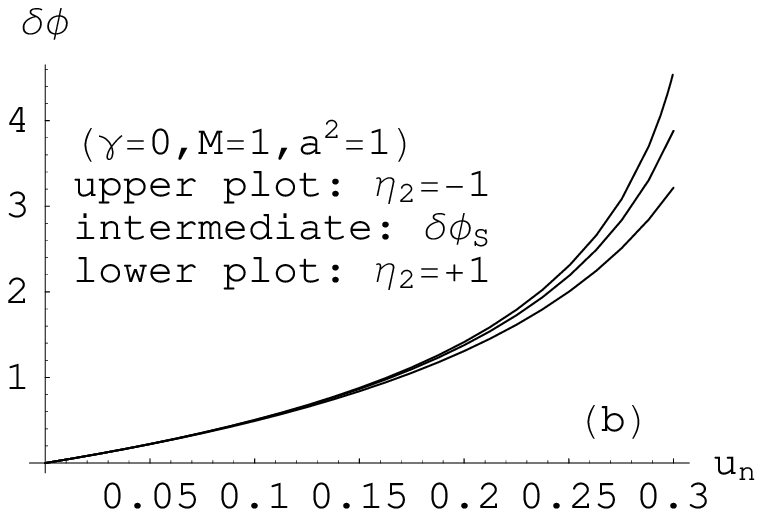}\\
  \caption{\footnotesize{The angle of deflection $\delta\phi$ (Eq.~\eqref{2.14}) in radians vs. $u_n=1/r_n$ ($r_n$ is the point on the null geodesic nearest the origin). In both plots the intermediate graph is the \s value $\delta\phi_{\text{S}}$. $\delta\phi$ ever increases and exceeds $2\pi$, then diverges, as $r_n$ decreases from $\infty$ to $r_{ps}$ (the photon sphere). $|\delta\phi-\delta\phi_{\text{S}}|$ decreases with $r$. (a): Phantom EMD cosh ($\eta_1=+1,\;\eta_2=-1$: upper plot) and normal EMD sinh ($\eta_1=+1,\;\eta_2=+1$: lower plot) \bh for $\ga=0>\ga_0$, $M=1$ and $a^2=1/4$. (b): Phantom EMD cosh ($\eta_1=+1,\;\eta_2=-1$: upper plot) and normal EMD sinh ($\eta_1=+1,\;\eta_2=+1$: lower plot) \bh for $\ga=0$, $M=1$ and $a^2=1$.}}\label{Fig1}
\end{figure}
\begin{figure}[h]
\centering
  \includegraphics[width=0.49\textwidth]{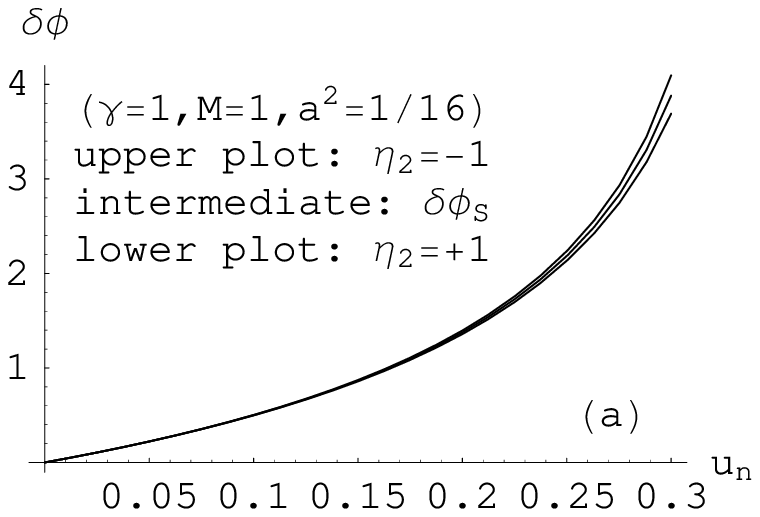}  \includegraphics[width=0.49\textwidth]{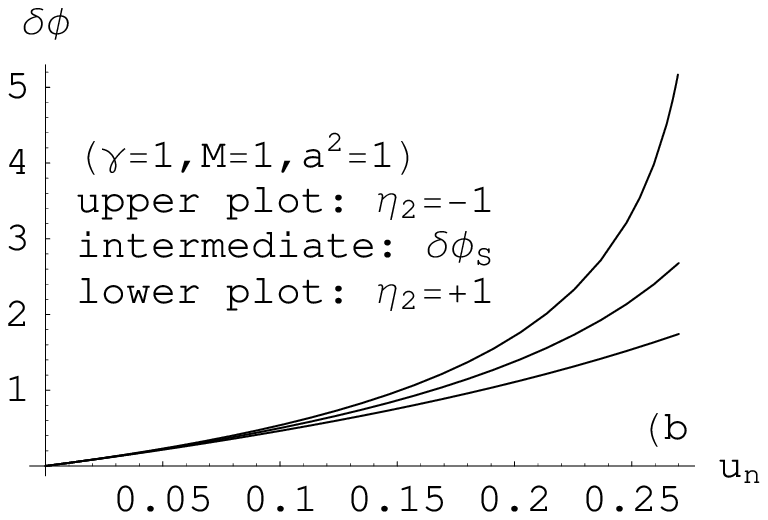}\\
  \caption{\footnotesize{The angle of deflection $\delta\phi$ (Eq.~\eqref{2.14}) in radians vs. $u_n=1/r_n$ ($r_n$ is the point on the null geodesic nearest the origin). In both plots the intermediate graph is the \s value $\delta\phi_{\text{S}}$. $\delta\phi$ ever increases and exceeds $2\pi$, then diverges, as $r_n$ decreases from $\infty$ to $r_{ps}$ (the photon sphere). $|\delta\phi-\delta\phi_{\text{S}}|$ decreases with $r$. (a): Phantom \RN ($\eta_2=-1$: upper plot) and normal \RN ($\eta_2=+1$: lower plot) \bh for $\ga=1>\ga_0$, $M=1$ and $a^2=1/16$. (b): Phantom \RN ($\eta_2=-1$: upper plot) and normal \RN ($\eta_2=+1$: lower plot) \bh for $\ga=1$, $M=1$ and $a^2=1$.}}\label{Fig2}
\end{figure}
\begin{figure}[h]
\centering
  \includegraphics[width=0.49\textwidth]{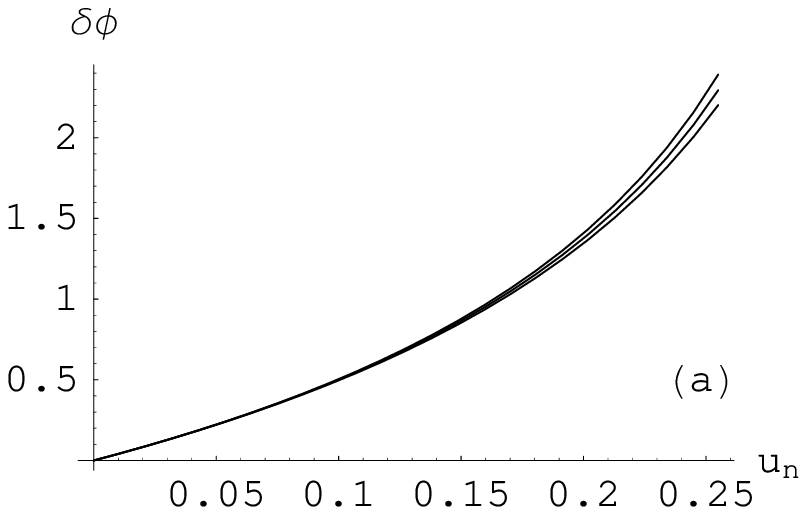}  \includegraphics[width=0.49\textwidth]{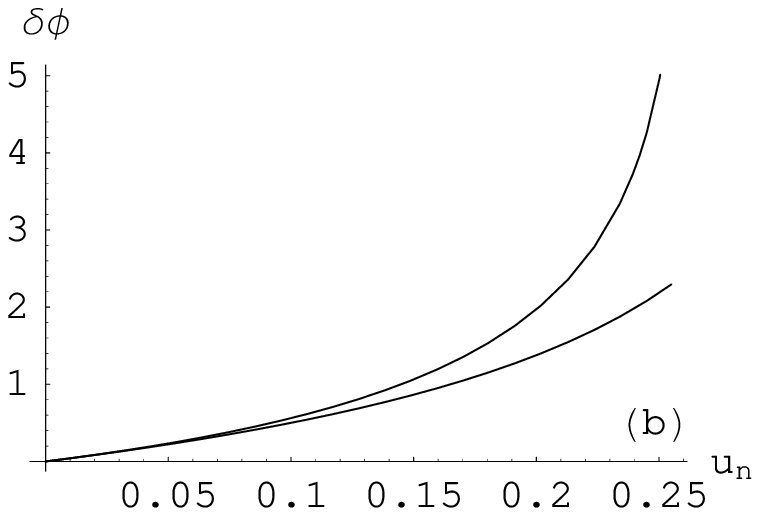}\\
  \caption{\footnotesize{The angle of deflection $\delta\phi$ (Eq.~\eqref{2.14}) in radians vs. $u_n=1/r_n$ ($r_n$ is the point on the null geodesic nearest the origin). $\delta\phi$ ever increases and exceeds $2\pi$, then diverges, as $r_n$ decreases from $\infty$ to $r_{ps}$ (the photon sphere). $|\delta\phi-\delta\phi_{\text{S}}|$ decreases with $r$. (a): E-anti-M-anti-D ($\eta_1=-1,\;\eta_2=-1$: upper plot), $\delta\phi_{\text{S}}$ for \s \abh (intermediate plot) and EM-anti-D ($\eta_1=-1,\;\eta_2=1$: lower plot) \bh for $\ga=10>\ga_0$, $M=1$ and $a^2=1/16$. (b): E-anti-M-anti-D ($\eta_1=-1,\;\eta_2=-1$: upper plot) and $\delta\phi_{\text{S}}$ for \s \abh (lower plot) for $\ga=10>\ga_0$, $M=1$ and $a^2=1$.}}\label{Fig3}
\end{figure}
\begin{figure}[h]
\centering
  \includegraphics[width=0.49\textwidth]{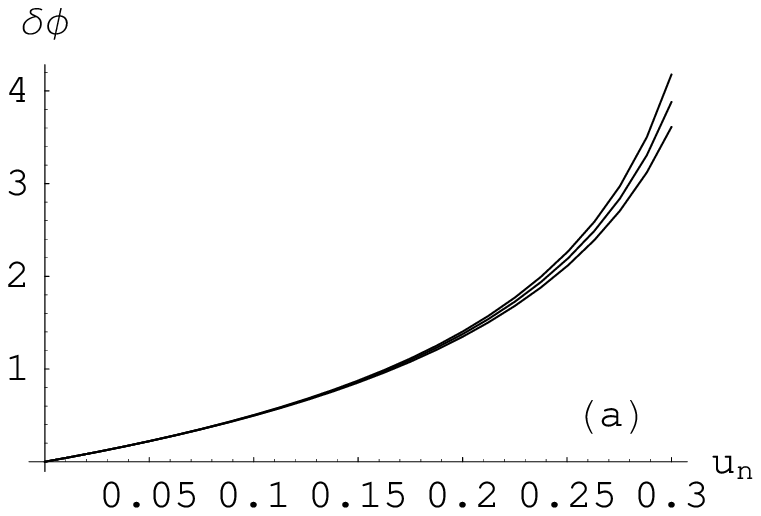}  \includegraphics[width=0.49\textwidth]{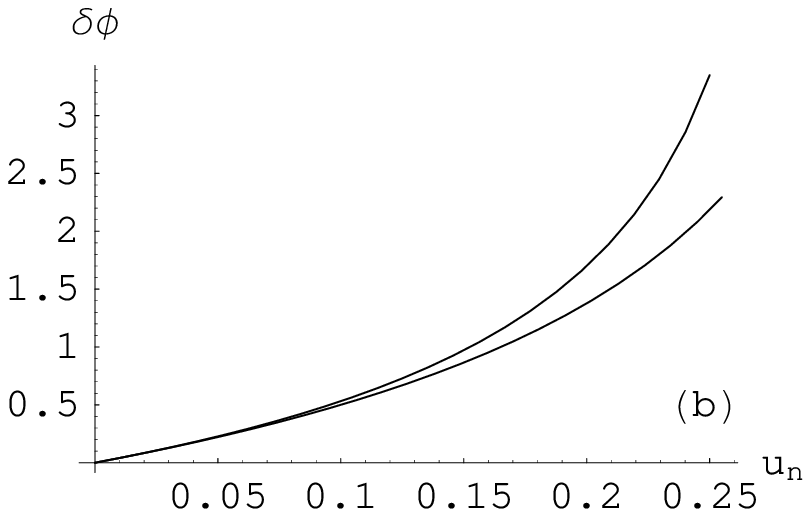}\\
  \caption{\footnotesize{The angle of deflection $\delta\phi$ (Eq.~\eqref{2.14}) in radians vs. $u_n=1/r_n$ ($r_n$ is the point on the null geodesic nearest the origin). In both plots the intermediate graph is the \s value $\delta\phi_{\text{S}}$. $\delta\phi$ ever increases and exceeds $2\pi$, then diverges, as $r_n$ decreases from $\infty$ to $r_{ps}$ (the photon sphere). $|\delta\phi-\delta\phi_{\text{S}}|$ decreases with $r$. (a): EMD ($\eta_1=1,\;\eta_2=1$: upper plot), $\delta\phi_{\text{S}}$ for \s \abh (intermediate plot) and E-anti-MD ($\eta_1=1,\;\eta_2=-1$: lower plot) \bh for $\ga=-1/2<\ga_0$, $M=1$ and $a^2=1/16$. (b): EMD ($\eta_1=1,\;\eta_2=1$: upper plot) and and $\delta\phi_{\text{S}}$ for \s \abh (lower plot) for $\ga=-1/2$, $M=1$ and $a^2=1$. These plots are different from those in Figures~(\ref{Fig1} to \ref{Fig3}) in that $\delta\phi$ for normal \bh exceeds that for phanton ones whenever the solution is defined.}}\label{Fig4}
\end{figure}

First consider the special case $\ga =0$, which corresponds to $\eta_1=+1$. In normal EMD ($\eta_2=+1$) we have $\delta\phi<\delta\phi_{\text{S}}$. In phantom EMD ($\eta_2=-1$), which is E-anti-MD theory, we have $\delta\phi>\delta\phi_{\text{S}}$ provided we restrict ourselves to the physical case $a^2<1$ [$4(3\pi-8)/\pi\simeq 1.8$]. Thus, in the presence of phantom fields, light rays are more deflected than in the normal case. Phantom fields cause light rays to bend with an angle $(3\pi-8)M^2a^2/(2r_n^2)$ larger than the angle of deflection caused by normal fields. This difference is independent of the sign of $q$ but depends on the mass of the black hole through $r_n$. Using~\eqref{2.16a} we obtain
\begin{equation}\label{2.19a}
    \frac{3\pi-8}{2r_n^2}\,M^2a^2\simeq \frac{3\pi-8}{32}\,a^2(\delta\phi_{\text{S}})^2
\end{equation}
which is $0.22 \%$ of the \s value $\delta\phi_{\text{S}}$ if $M=1$, $a^2=1/4$, $u_n=0.05$ and $0.89 \%$ of it (the exact value is $1.03 \%$) if $M=1$, $a^2=1$, $u_n=0.05$.

The case $\ga =1$ is phantom or normal \RN black hole. With
\begin{equation}\label{2.20}
    \delta\phi = \delta\phi_{\text{S}}-\eta_2 \frac{3\pi M^2a^2}{4}\,\frac{1}{r_n^2}+\cdots\,,\quad \text{if }\ga =1
\end{equation}
we confirm the previous conclusions: $\delta\phi<\delta\phi_{\text{S}}$ for normal Reissner-Nordstr\"om \bh and $\delta\phi>\delta\phi_{\text{S}}$ for phantom ones. A phantom Reissner-Nordstr\"om black hole deflects light with an angle $3\pi q^2/(2r_n^2)$ larger than the deflection angle caused by a normal Reissner-Nordstr\"om black hole
\begin{equation}\label{2.19b}
    \frac{3\pi}{2r_n^2}\,M^2a^2\simeq \frac{3\pi}{32}\,a^2(\delta\phi_{\text{S}})^2
\end{equation}
which is $1.47 \%$ of the \s value $\delta\phi_{\text{S}}$ if $M=1$, $a^2=1/4$, $u_n=0.05$ and $5.89 \%$ of it (the exact value is $6.56 \%$) if $M=1$, $a^2=1$, $u_n=0.05$.

Now, consider the case $\ga \neq 0$ and $\ga \neq 1$ ($\ga \neq -1$). Here again we confirm the previous conclusions: $\delta\phi<\delta\phi_{\text{S}}$ for normal \bh and $\delta\phi>\delta\phi_{\text{S}}$ for phantom ones provided $|\ga|$ is large enough. This is no longer true if $\ga$ is closer to $-1$ as the coefficient of $1/r_n^2$ becomes too large invalidating~\eqref{2.19}. The relative discrepancy function $R_{\de\phi}=(\de\phi-\delta\phi_{\text{S}})/\delta\phi_{\text{S}}$ is shown in Figures~\ref{Fig5} and~\ref{Fig6}, which have been plotted using the exact formula~\eqref{2.14}. The Figures illustrate the existence of a zero $\ga_0$ beyond which $R_{\de\phi}>0$  ($\delta\phi>\delta\phi_{\text{S}}$) for phantom \bh and  $R_{\de\phi}<0$ ($\delta\phi<\delta\phi_{\text{S}}$) for normal ones.

Figures~\ref{Fig5} and~\ref{Fig6} have been plotted for fixed ($M=1,a^2=1/16$) and a relatively large value of $r_n$ ($u_n=0.05$). Based on these and on some other figures (not shown in this paper) for small values of $r_n$ up to the photon sphere ($u_n=0.3$) and larger values of $a^2$ up to 1, we can draw a general conclusion: For fixed ($M,a^2,u_n$), there is always a root $\ga_0$ in the interval ($-0.2,0$) to $R_{\de\phi}(\ga)=0$. Otherwise, for some values of $\ga$ in the interval ($-0.2,0$), it seems there is always a critical value $r_n=r_c$, larger than the photon sphere, where $R_{\de\phi}=0$.

As $\ga\to +\infty$, $R_{\de\phi}$ approaches the limit
\begin{align*}
& \frac{2}{\delta\phi_{\text{S}}}\int_{0}^{1}\frac{\e ^{K_-x}\,\dd x}{\sqrt{\e ^{K_-}(1-K_+)-x^2\e ^{K_-x}(1-K_+x)}}-\frac{\pi+\delta\phi_{\text{S}}}{\delta\phi_{\text{S}}}\\
&K_{\pm}=Mu_n(\sqrt{1-2\eta_2a^2}\pm 1)\,.
\end{align*}

As Figures~\ref{Fig1} to~\ref{Fig4} reveal, the vertical spacing $|\delta\phi-\delta\phi_{\text{S}}|$, whenever defined, depends slightly on $\ga$, which itself depends on $\eta_1$, and increases with $a^2$. In the extreme case ($a^2=1$), the winding number for phantom \bh with $\eta_2=-1$ and $\ga > \ga_0$ (regardless of the sign of $\eta_1$) diverges near $u_n\simeq 0.3$, a value for which the angle of deflection for phantom ($\eta_1=-1$) or normal ($\eta_1=+1$) \bh with $\eta_2=+1$ is less than a few radians. As we shall see later, this is a consequence of the fact that the photon sphere for \bh with $\eta_2=-1$ (\bh where the Maxwell field $F$ is coupled repulsively to gravity) and $\ga > \ga_0$ is larger than $3M$, which is the \s limit, allowing photons to orbit the hole at larger, ever-decreasing, radii. The \s limit $3M$ is larger than the photon sphere for \bh with $\eta_2=+1$ and $\ga > \ga_0$. If $\ga<\ga_0$ and $\eta_2\ga R_{\de\phi}<0$, all that is true for \bh with $\eta_2=-1$ (respectively, $\eta_2=+1$) applies to \bh with $\eta_2=+1$ (respectively, $\eta_2=-1$).

It is useful to express $\delta\phi$ in terms of the charges ($M,q$) and the impact parameter $b=L/E$. For that end we need to solve the equation $E^2=L^2g(u_n)$ (see Appendix A) for $u_n=1/r_n$ to the second order in $1/b$. This equation is equivalent to $b=r_n/\sqrt{f_+(u_n)f_{-}(u_n)^{2\gamma-1}}$. We obtain
\begin{equation*}
    u_n=\frac{1}{b}+\frac{r_++r_-(2\ga -1)}{2}\,\frac{1}{b^2}+O[1/b]^3\,.
\end{equation*}
Using this in~\eqref{2.16} we derive the desired equation
\begin{equation}\label{2.21}
    \delta\phi = \frac{4M}{b}+ \frac{\pi }{16} [15r_+^2+6 r_-r_+(4 \gamma -1)+r_-^2(16\gamma ^2-1)]\,\frac{1}{b^2}+O[1/b]^3\,.
\end{equation}
In terms of the \s value, $\delta\phi_{\text{S}}=(4M/b)+[15\pi M^2/(4b^2)]+O[1/b]^3$, $b$ and ($M,q$), the expressions~(\ref{2.17}, \ref{2.18}, \ref{2.20}) become, respectively,
\begin{align}
\label{2.23}&\delta\phi = \delta\phi_{\text{S}}-\frac{\pi q^2}{16}\big[12\eta_2+\frac{q^2}{M^2}\big]\,\frac{1}{b^2}+\cdots\,,\; \text{if }\ga =0\\
\label{2.24}&\delta\phi = \delta\phi_{\text{S}}-\eta_2\big[\frac{(\ga -1)[16\ga -\pi(\ga +1)]M|M-\mathcal{M}|+ [\pi\ga (7\ga -1)+16\ga^2]q^2}{8\ga^2}\big]\,\frac{1}{b^2}+\cdots\,,\; \forall \ga \neq -1\\
\label{2.25}&\delta\phi = \delta\phi_{\text{S}}-\eta_2 \frac{(3\pi +8) q^2}{4}\,\frac{1}{b^2}+\cdots\,,\; \text{if }\ga =1\,.
\end{align}

\paragraph*{\textbf{Time delay.}}

We evaluate the coordinate time $T(U)$ required for light to travel from a point $U=1/R$ to $u_n=1/r_n$ in the plane $\ta=\pi/2$. Using Eqs~(\ref{2.10}, \ref{2.12}) along with $E^2=L^2g(u_n)$, we obtain
\begin{equation}\label{d1}
    T(U)=\int_{U}^{u_n}\frac{\sqrt{g(u_n)}\,\dd u}{u^2f_+(u)f_-(u)^{\ga}\sqrt{g(u_n)-g(u)}}=\frac{1}{u_n}\int_{X}^{1}\frac{\sqrt{g(u_n)}\,
    \dd x}{x^2f_+(u_nx)f_-(u_nx)^{\ga}\sqrt{g(u_n)-g(u_nx)}}
\end{equation}
where we have set $u=u_nx$ and $X=U/u_n=r_n/R$ satisfies $0<X<1$. An expansion in terms of powers of $u_n$ leads to
\begin{align}
\label{d2}&T(U)=T_{\text{S}}+\frac{ \sqrt{1-X^2}}{ (1+X)} \frac{(\gamma -1) \eta _2 M^2 a^2}{(\gamma +1) (M+\mathcal{M})} +O[u_n]^2\;\text{ if }\ga \neq -1\\
\label{d3}&T(U)=T_{\text{S}}-\frac{3\eta _2 M^2 a^2}{2}\Big(\frac{\pi}{2}-\arcsin X\Big)u_n+O[u_n]^3\;\text{ if }\ga = 1
\end{align}
where $T_{\text{S}}$ is the corresponding \s value~\cite{del1}
\begin{multline*}
    T_{\text{S}}=\sqrt{R^2-r_n^2}+\frac{M \sqrt{1-X^2}}{1+X}+2 M \ln \big[\frac{1+\sqrt{1-X^2}}{X}\big]\\+M^2 \Big[\frac{15}{2}
    \Big(\frac{\pi }{2}-\arcsin  X\Big)-\frac{(4+5X) \sqrt{1-X^2}}{2 (1+X)^2}\Big] u_n+O[u_n]^3\,.
\end{multline*}
Notice that the correction, which has been added to the \s value in~\eqref{d2}, vanishes for \RN \bh ($\ga=1$), and a second order correction is needed for these \bh as shown in~\eqref{d3}. Since the sign of $(\ga-1)/(\ga+1)$ is, by~\eqref{2.4}, that of $-\eta_1$, we conclude from~\eqref{d2} that $T(U)>T_{\text{S}}$ if $\eta_1\eta_2<0$ (E-anti-MD or ED-anti-M) and that $T(U)<T_{\text{S}}$ if $\eta_1\eta_2>0$ (EMD or E-anti-D-anti-M). For \RN \bh the time $T(U)$ is such that $T(U)>T_{\text{S}}$ for E-anti-M and $T(U)<T_{\text{S}}$ for EM.

The time delay $\De T$ or the excess in time is the difference between the time required for light to travel from the source located at the point $U_s=1/R_s$ through the point $u_n=1/r_n$ to the observer located at the point $U_o=1/R_o$ and the time needed for light to travel from $U_s$ to $U_o$ in the absence of gravity (and any field that may cause light to deflect). As shown in~\cite{del2} the time delay may be positive, zero or negative.

Using the same notation as in Figures 1 of~\cite{images,11a}, the axis joining the observer $O$ and the \abh (the lens or deflector $L$) is the optic axis. The angle $\widehat{LOS}$ is $\bt$ and the angle $\widehat{LOI}$ is $\ta$ where $I$ is the image of the source $S$ on the same side as the latter. $D_s$ and $D_d$ are the distances of the source and the lens from the observer (measured along the optic axis) and $D_{ds}=D_s-D_d$ represents the projected distance on the optic axis from the lens to the source. With these notations we have
\begin{equation*}
\De T=T(U_s)+T(U_o)-D_s\sec \bt\,.
\end{equation*}
Far away from the lens, for large values of the impact parameter, one may expand $\De T$ in powers of $\epsilon=\ta_E/(4D)$ where $\ta_E$ is a good estimate of the angular radius of the Einstein ring of \s lensing
\begin{equation*}
\ta_E^2=4M\,\frac{D_{ds}}{D_dD_s}\,,\quad D=\frac{D_{ds}}{D_s}\,.
\end{equation*}
For $\ga \neq -1$ we have
\begin{equation}\label{d4}
\De T=2 M \Big\{1+\frac{\beta ^2-\theta ^2}{\theta_E^2}-\ln \big[\frac{D_d \theta ^2}{4 D_{ds}}\big]
+\frac{(\gamma -1) \eta_2 M a^2}{(\gamma +1) (M+\mathcal{M})}\Big\}+O[\epsilon]^3\,,\;(\ga \neq -1)
\end{equation}
[with $2M=(8D_dD_{ds}/D_s)\epsilon^2$], where the terms independent of $\ga$ correspond to the \s value $\De T_S$. For $\ga=1$, the last term in~\eqref{d4} vanishes and a term proportional to $\epsilon^3$ is needed. Expanding $\ta=\ta_0+\ta_1\epsilon+O[\epsilon]^2$, we obtain\footnote{In~\eqref{d4} we only needed $\ta=\ta_0+O[\epsilon]$.}
\begin{equation}\label{d5}
\De T=\De T_S-6\pi\frac{D_dD_{ds}}{D_s}\frac{\eta_2a^2\ta_E}{\ta_0}\epsilon^3+O[\epsilon]^4\,,\;(\ga = 1)
\end{equation}
where $\De T_S$ is the \s value up to $\epsilon^3$ as derived in~\cite{del1}.

\setcounter{equation}{0}
\section{Determination of geodesics by \W: general procedure}\label{3}

Any differential equation of the form~\eqref{1.2} has a unique solution in terms of the \aw of the form~\cite{Wang,Whittaker}
\begin{equation*}
    \ro = \wp(\Ta+C)
\end{equation*}
where $C$ is generally a complex constant.

$\wp(\Ta)$ is an even single-valued doubly periodic function with half periods ($\om,\om'$) chosen in such a way that ${\rm Im}(\om'/\om)>0$. When the \ws polynomial $w(\ro)=4\ro^3-g_2\ro-g_3=4(\ro-e_1)(\ro-e_2)(\ro-e_3)$ has three real roots ($e_3,e_2,e_1$), there are three half periods ($\om_1,\om_2,\om_3$) depending on ($\om,\om'$) such that
\begin{equation}\label{3.1}
    \wp(\om_k)=e_k\,,\quad (k=1,2,3)\,.
\end{equation}
To have $e_3<e_2<e_1$ we choose the three half periods ($\om_1,\om_2,\om_3$) to satisfy~\cite{Wang}
\begin{equation}\label{3.2}
    \om_1\equiv \om>0\,,\quad \om_3\equiv \om'\;(\text{with } -\ii \om'>0)\,,\quad \om_2\equiv -\om-\om'=-\om_1-\om_3\,.
\end{equation}
The expression of $\om_2$ is a consequence of $e_1+e_2+e_3=0$.

\subsection{Three distinct real roots}

The \ws polynomial $w(\ro)$ will have three real roots if
\begin{equation}\label{3.3}
    g_2>0 \;\text{ and }\; \De\equiv g_2^3-27g_3^2>0\,.
\end{equation}
We parametrize the (real) roots by the angle $0\leq \eta\leq \pi$ as follows:
\begin{align}\label{3.4}
&     e_3=-\sqrt{\frac{g_2}{3}}\cos\Big(\frac{\pi-\eta}{3}\Big)<0\,,\;\,
    e_2=-\sqrt{\frac{g_2}{3}}\cos\Big(\frac{\pi+\eta}{3}\Big)\,,\;\,
    e_1=\sqrt{\frac{g_2}{3}}\cos\Big(\frac{\eta}{3}\Big)>0\\
&\cos\eta =\frac{9g_3}{\sqrt{3g_2^3}}\,,\quad \sin\eta =\sqrt{\frac{\De}{g_2^3}}>0\,.\nn
\end{align}
The signs of $e_3<0$, $e_1>0$, and $\sin\eta>0$ are well defined,  $e_3<e_2<e_1$, and the sign of $e_2$ depends on $g_3$:
\begin{equation}\label{3.5}
    e_2\cdot g_3<0\;\text{ and }\;e_2=0\text{ if }g_3=0\qquad (g_3=4e_1e_2e_3)\,.
\end{equation}
Motion is possible where $w(\ro)\geq 0$:
\begin{equation}\label{3.6}
    e_3\leq\ro\leq e_2\;\text{ or }\;\ro\geq e_1\,.
\end{equation}

Conversely, we can reverse~\eqref{3.1} and express the half periods ($\om_1,\om_2,\om_3$) in terms of the roots\footnote{There is a third expression for $\om_3$ that appears with a misprinted sign in~\cite{Whittaker,GV}. The correct expression is $\om_3=+\ii \int_{-\infty}^{e_3}\frac{\dd \ro}{\sqrt{-w(\ro)}}$.}~\cite{Wang}.
\begin{align}
\label{3.7}&\om_1=\int_{e_1}^\infty\frac{\dd \ro}{\sqrt{w(\ro)}}=\int_{e_3}^{e_2}\frac{\dd \ro}{\sqrt{w(\ro)}} \\
\label{3.8}&\om_3=\int_{e_3}^\infty\frac{\dd \ro}{\sqrt{w(\ro)}}= \ii \int_{e_2}^{e_1}\frac{\dd \ro}{\sqrt{-w(\ro)}}\,.
\end{align}

Let $\ri,\;\ro_0$ be the values of $\ro$ corresponding to $r=+\infty,\;r=0$, respectively, and let $\ro_+,\;\ro_-$ correspond to $r=r_+,\;r=r_-$, if there are any\footnote{In case of wormhole solutions, one introduces $\ro_a$ corresponding to the radius $a$ of the throat.}. A singularity is denoted by $\ro_{\text{sing}}$ ($r_{\text{sing}}$), which may be any of $\ro_0$, $\ro_-$ depending on the theory. In a general physical situation ($\ri,\ro_0,\ro_+,\ro_-,\ldots,e_3,e_2,e_1$) are functions\footnote{Some of which may be constants as in the case of \s solution where $e_3<\ri=-1/12<e_2$ and $\ro_0=+\infty$.} of the vector of parameters $\vec{p}$ $=$ (charges, constants of motion) $=$ ($M,q,\ldots ,E,L,\ldots$) so that the locations of these points on the $\ro$-axis change with $\vec{p}$. We shall represent ($e_3,e_2,e_1$) at the same locations on the $\ro$-axis while ($\ri,\ro_0,\ro_+,\ro_-,\ldots$) appear on different locations depending on $\vec{p}$.

To determine all types of geodesic motion in a given geometry, one has to consider all allowed possible locations of ($\ri,\ro_0,\ro_+,\ro_-,\ldots$) with respect to ($e_3,e_2,e_1$). Once this is done, any geodesic motion that can be brought to~\eqref{1.2} is integrated by mere comparison with the work done in this section. We shall provide some examples in this section and in the next two we apply the procedure to phantom and normal \RN and EMD \BH. To illustrate the procedure, we shall envisage only some locations of ($\ri,\ro_0,\ro_+,\ro_-,\ldots$) with respect to ($e_3,e_2,e_1$), most of which are related to phantom and normal \RN and EMD \BH.

\subsubsection{Scattering and trapped paths}\label{3.1.1}
We consider four possible situations.

\paragraph{$\pmb{e_3<\ri<e_2<e_1<\ro_0=\ro_{\text{sing}}}$.} As light scatters from $r=+\infty\to r_n\to r=+\infty$, the corresponding point on the $\ro$-axis moves from $\ri\to e_2\to\ri$, if $\ro(r)$ is a decreasing function of $r$, or from $\ri\to e_3\to\ri$, if $\ro(r)$ is an increasing function of $r$. We consider the former case throughout this section, which is also going to be the case for the next two sections. The solution to~\eqref{1.2} is
\begin{equation*}
    \ro(r)=\wp(\Ta(\phi)+C)\,.
\end{equation*}
To fix $C$ we may assume $\Ta=0$ at $\ro_n=e_2$, corresponding to $r_n$, or assume $\Ta=0$ at $\ri$. The former case looks simpler leading to $e_2=\wp(C)$ and thus, by~\eqref{3.1}, $C=\om_2$ or $C=-\om_2$ ($\wp$ is an even function!). We choose the latter solution so that by~\eqref{3.2} $C=\om_1+\om_3$ and
\begin{equation}\label{3.9}
    \ro(r)=\wp(\Ta(\phi)+\om_1+\om_3)\,.
\end{equation}

The angles $\Ta$ and $\phi$ are related by a linear formula that may be put on the form $\Ta=\kappa \phi$. The angle of deflection is then given by
\begin{equation}\label{3.9b}
    \de\phi=\frac{2}{|\kappa|}\int_{\ri}^{e_2}\frac{\dd \ro}{\sqrt{w(\ro)}}-\pi=\frac{2}{|\kappa|}\bigg(\int_{e_3}^{e_2}-\int_{e_3}^{\ri}\bigg)\frac{\dd \ro}{\sqrt{w(\ro)}}-\pi=\frac{2}{|\kappa|}\om_1-\frac{2}{|\kappa|}\int_{e_3}^{\ri}\frac{\dd \ro}{\sqrt{w(\ro)}}-\pi
\end{equation}
where we have used the second formula in~\eqref{3.7}.

The other possible motion, the so-called trapped or terminating bound path, is in the region between $e_1$ and $\ro_0$ where $w(\ro)\geq0$. If the path starts from the singularity $\ro_0=\ro_{\text{sing}}$ ($r=0$), it reaches the farthest point $\ro_f=e_1$ ($r=r_f$) and then returns to $\ro_0$. Again choosing $\Ta=0$ at the farthest point, the solution is
\begin{equation}\label{3.10}
    \ro(r)=\wp(\Ta(\phi)+C)\,,\;\text{ with }\;C=\om_1\;\text{ (or }\;C=-\om_1)\,.
\end{equation}

\paragraph{$\pmb{e_3<\ri<e_2<e_1<\ro_+<\ro_0<\ro_-=+\infty}$.} If $\ro_0$ is a singularity, then this case is identical to (a) with a scattering path from $\ri\to e_2\to\ri$ given by~\eqref{3.9} and a trapped path between $e_1$ and $\ro_0$ given by~\eqref{3.10}.

If  $\ro_-$ is a singularity but $\ro_0$ is not, then there is a trapped path between $e_1$ and $\ro_-$ given by~\eqref{3.10}.

If neither $\ro_0$ nor $\ro_-$ is a singularity, the path is a many-world periodic bound orbit~\cite{met} in that the path, after crossing the inner horizon at $r=r_-$, emerges in another copy of the space-time then in another copy of it and so on. If we choose $\Ta =0$ at $\ro=e_1$, then the solution will be given by~\eqref{3.10}.

\paragraph{$\pmb{e_2<\ro_0<e_1<\ri<\ro_+<\ro_-=+\infty}$.} Since $w(\ro)<0$ for $\ro\in (e_2,e_1)$, there are no paths that can reach or emanate from the origin.

There is a path that extends from spatial infinity ($\ri$) to the inner horizon ($\ro_-$). This is not a spiral path since the integral
\begin{equation*}
\int_{\text{const}\geq\ri}^{\infty}\dd \ro/\sqrt{w(\ro)}
\end{equation*}
converges ($\Ta$ remains finite). If $\ro_-$ is not a singularity, the path is called a two-world scattering orbit in that the path emerges, after crossing the inner horizon at $r=r_-$, in another copy of the space-time and back to spatial infinity. If $\ro_-$ is a singularity, we have an absorbed path from spatial infinity to the singularity. The solution is again $\ro(r)=\wp(\Ta(\phi)+C)$. Since there is no farthest or nearest point on the path, we choose $\Ta =0$ at $\ri$ leading to $\ri =\wp(C)$. Using the inverse function to $\wp$, $C=\int_{\ri}^{\infty}\dd \ro/\sqrt{w(\ro)}$ and
\begin{equation}\label{3.11}
    \ro(r)=\wp(\Ta(\phi)+C)\,,\;\text{ with }\;C=\int_{\ri}^{\infty}\dd \ro/\sqrt{w(\ro)}\,.
\end{equation}

\paragraph{$\pmb{\ro_0<e_3<\ri<e_2<e_1<\ro_-=+\infty}$.} Here again no paths that can reach or emanate from the origin.

We have a scattering path from $\ri\to e_2\to\ri$, and the solution is again given by~\eqref{3.9} but with different $\vec{p}$.

There is another path from $r=r_1$ ($\ro=e_1$) to $r=r_-$ ($\ro=\ro_-$). Since in this case $r_1$ is finite ($\ri<e_1$ $\Rightarrow$  $r_1<+\infty$), the path is a many-world periodic bound orbit if $\ro_-$ is not a singularity or a trapped path if $\ro_-$ is a singularity. If we choose $\Ta =0$ at $\ro=e_1$, then the solution will be given by~\eqref{3.10}.

\subsubsection{Absorbed and circular paths}
Absorbed paths extend from spatial infinity ($\ri$) to the (nearest)  singularity ($\ro_{\text{sing}}$). Such paths exist if both points $\ri,\;\ro_{\text{sing}}$ are in $[e_3,e_2]$ or in $[e_1,+\infty)$. [There are no such paths in the \s case when $w(\ro)$ has three distinct real roots.] It is clear that there are no circular paths when $w(\ro)$ has three distinct real roots.

\subsection{Two distinct real roots}\label{3b}

The \ws polynomial $w(\ro)$ will have two real roots if
\begin{equation}\label{3.12}
    g_2>0 \;\text{ and }\; \De\equiv g_2^3-27g_3^2=0\,.
\end{equation}
This happens when one of the local extreme values of $w(\rho)$ is zero. The second condition in~\eqref{3.12} splits  into two cases.

\subsubsection{Stable circular and bound paths: $\pmb{g_3=(g_2/3)\sqrt{g_2/3}>0}$}

The local maximum value of $w(\rho)$, which is at $\ro_{\text{max}}=-(1/2)\sqrt{g_2/3}$, is zero. We have
\begin{equation}\label{3.13}
    e_3=e_2=-\frac{1}{2}\sqrt{\frac{g_2}{3}}\,,\quad e_1=\sqrt{\frac{g_2}{3}}\quad [\eta =0 \text{ in}~\eqref{3.4}]\,.
\end{equation}
Since at $\ro_{\text{max}}=-(1/2)\sqrt{g_2/3}$, $w(\rho)$ has a local maximum, the polynomial $P(r)$ in~\eqref{1.1}
has a local maximum too at the corresponding point $r_{\text{max}}$. But since $P(r)\propto E^2-V(r)$ [compare with~(\ref{2.12})], the potential $V(r)$ has there a local minimum. Thus
\begin{equation}\label{3.14}
    \ro\equiv \ro_{\text{max}}=e_3= -(1/2)\sqrt{g_2/3}
\end{equation}
is a stable circular path.

Paths in the region $\ro \geq e_1$ are periodic: They include the periodic bound and the so-called terminating bound (trapped) and many-world periodic bound orbits~\cite{met}. No matter the location of $\ri$ with respect to $e_1$ is, the equation of motion can be integrated directly. Let $t=\sqrt{\ro-e_1}$ and $k^2=e_1-e_3=(3/2)\sqrt{g_2/3}>0$, and then~\eqref{1.2} reads
\begin{equation*}
    \dd \Ta = \frac{\dd \ro}{2(\ro-e_3)\sqrt{\ro-e_1}}=\frac{\dd t}{t^2+k^2}
\end{equation*}
leading to
\begin{equation}\label{3.16}
    \sqrt{\ro-e_1}=k\tan[k(\Ta -C)] \;\; \text{ or }\;\; \ro-e_3=\frac{2(e_1-e_3)}{1+\cos[2k(\Ta -C)]}\,.
\end{equation}
For the \s \abh $\ro =M/(2r)-1/12$~\cite{GV}, $\Ta =\phi$, $g_2=1/12$, $g_3=1/216$, $e_3=e_2=-1/12$, $e_1=1/6$, $k=1/2$, and the second formula in~\eqref{3.16} is just Eq.~(10) of~\cite{GV}.

For the phantom ($\eta_2=-1$) and normal ($\eta_2=+1$) \RN \bh we derive in Section~\ref{5.two} from the first formula in~\eqref{3.16} the following orbit
\begin{equation}\label{3.17}
    \tan\Big[\frac{\phi -C}{2}\Big]=\sqrt{\frac{r_+-r}{r-r_-}}
\end{equation}
which is a trapped path for the phantom \abh ($r_+>r>0$) and a many-world periodic bound path for the normal \abh ($r_+>r>r_-$). Using the double-angle formula for tan we rewrite it as
\begin{equation}\label{3.18}
    \tan(\phi -C) =\frac{M-r}{\sqrt{2Mr-\eta_2 q^2-r^2}}\,,
\end{equation}
which is the correct form of the misprinted Eq.~(64) of~\cite{GV}.

\subsubsection{Unstable circular and spiral paths: $\pmb{g_3=-(g_2/3)\sqrt{g_2/3}<0}$}

The local minimum value of $w(\rho)$, which is at $\ro_{\text{min}}=+(1/2)\sqrt{g_2/3}$, is zero. We have
\begin{equation}\label{3.19}
    e_2=e_1=+\frac{1}{2}\sqrt{\frac{g_2}{3}}\,,\quad e_3=-\sqrt{\frac{g_2}{3}}\quad [\eta =\pi \text{ in}~\eqref{3.4}]\,.
\end{equation}
Since at $\ro_{\text{min}}=+(1/2)\sqrt{g_2/3}$, $w(\rho)$ has a local minimum, the potential $V(r)$ has there a local maximum [compare with~(\ref{2.12})]. Thus
\begin{equation}\label{3.20}
    \ro\equiv \ro_{\text{min}}=e_1= +(1/2)\sqrt{g_2/3}
\end{equation}
is an unstable circular path.

Paths in the regions $e_3\leq\ro\leq e_2$ and $\ro \geq e_1$ depend on the location of $\ri$. Here we consider the case $e_3<\ri < e_2$. There are two spiral paths approaching the circle $\ro=e_1=+(1/2)\sqrt{g_2/3}$ from 1) $\ri$ or from 2) $\ro_0$ ($\ro_0>e_1$ in this case). The paths end orbiting the center at an ever 1) decreasing or 2) increasing radii $r$ without, however, reaching the unstable circular path at $r=r_1$ corresponding to $\ro=e_1$.  The equation of motion can be integrated directly. Let $s=\sqrt{\ro-e_3}$ and $k^2=e_1-e_3=(3/2)\sqrt{g_2/3}>0$. Then~\eqref{1.2} reads
\begin{equation*}
    \dd \Ta = \frac{\dd \ro}{2|\ro-e_1|\sqrt{\ro-e_3}}=\frac{\dd s}{|s^2-k^2|}
\end{equation*}
and $\Ta$, as well as the angle of deflection, diverge as $\ln |\ro-e_1|$ as $\ro\to e_1$, which is a general behavior in the strong field limit valid for all spherically symmetric solutions~\cite{11a,11,13}. Integration leads to
\begin{equation}\label{3.21}
    \text{1) }\sqrt{\ro-e_3}=-k\coth[k(\Ta -C)] \;\; \text{ and }\;\; \text{2) }\sqrt{\ro-e_3}=-k\tanh[k(\Ta -C)]
\end{equation}
or
\begin{equation}\label{3.22}
    \ro-e_1=-\frac{2(e_1-e_3)}{1\mp \cosh[2k(\Ta -C)]}\,,\quad [-\to\text{ 1)},\;+\to\text{ 2)}]\,.
\end{equation}
For the \s \abh $\ro =M/(2r)-1/12$~\cite{GV}, $\Ta =\phi$, $g_2=1/12$, $g_3=-1/216$, $e_3=-1/6$, $e_1=e_2=1/12$ and $k=1/2$ we obtain the solutions~(11) of~\cite{GV}.

\subsection{One real root}

The \ws polynomial $w(\ro)$ will have one real root with multiplicity 1 if
\begin{equation}\label{3.23}
    \De\equiv g_2^3-27g_3^2<0\,.
\end{equation}
The sign of the real root $e_r$
\begin{equation}\label{4.23c}
    e_r=\frac{1}{2\cdot 9^{1/3}}[(9 g_3+ \sqrt{3} \sqrt{-\Delta })^{1/3}+(9 g_3-\sqrt{3} \sqrt{-\Delta })^{1/3}]
\end{equation}
is related to that of $g_3$ by
\begin{equation}\label{3.24}
    e_r\cdot g_3>0\;\text{ and }\;e_r=0\text{ if }g_3=0\,.
\end{equation}
Motion is possible for $\ro\geq e_r$.

Absorbed paths exist if the range $e_r\leq\ro<\infty$ includes $r_{\text{sing}}<r<\infty$. In that case, we will have $e_r\leq\ri<\ro_{\text{sing}}$ and the solution will be given by~\eqref{3.11} where the upper limit of integration ``$\infty$" is replaced by ``$\ro_{\text{sing}}$".

If $\ri<e_r<\ro_{\text{sing}}$, the solution is a trapped path given by~\eqref{3.10}.

One can envisage other situations as $\ro_{\text{sing}}<e_r$ and so on. However, we are giving examples that are more or less related to EMD and \RN \BH.

The case $g_2=g_3=0$ implies $e_r=0$ [this is the only case where the three real roots of $w(\ro)=0$ are equal]. This is no different from the two cases discussed above for the generic case. However, if $\ro(r)$ were an increasing function of $r$  and $\ri>e_r=0$, the angle $\Ta$ would diverge as
\begin{equation*}
    \Ta -C=\int_{\ro}^{e_r=0}\frac{\dd \ro'}{\sqrt{4\ro'^3}}\propto \lim_{\ro'\to 0^{+}}\frac{1}{\sqrt{\ro'}}
\end{equation*}
for paths approaching $e_r=0$ from the right. This is not a logarithmic behavior as the one we have seen earlier. These spiral paths would approach, without reaching, the unstable circular path at $\ro=e_r=0$.

\subsection{Application: Strong field limit and relativistic images}\label{3d}

Because of the orbiting effect~\cite{Az,orb} geodesics may orbit many times around the deflector before escaping to infinity. In the case of light paths, images that are formed because of geodesic deflection by more than $3\pi/2$ are called relativistic images~\cite{images}. As we have seen in Subsection~\ref{3b}, such an orbiting effect happens in the strong field region and takes place as the nearest point $r_n$ (to the origin) approaches (but remains larger than) the radius of the photon sphere $r_{ps}$. The limit $r_n\to r_{ps}$ leads to spiral paths approaching endlessly the photon sphere. This provides the most common definition of the photon sphere~\cite{ps1}\footnote{An alternative equivalent definition of the photon sphere has been formulated in~\cite{ps2} along with an energy condition for a \aBH, in a static spherically symmetric spacetime, to be surrounded by it.}. In the context of \W, we have established that photon spheres are the unstable circular paths with two equal positive roots of the \ws polynomial, $e_1=e_2>0$, and a negative root, $e_3=-2e_1$.

In the strong field limit, all relations governing the light trajectory as well as the determination of the angular positions of images and related entities form a set of transcendental, analytically nontractable, equations and inequations. Numerical solutions to the case of a \s \abh exit and have led to the following conclusions~\cite{ri}: 1) Observations of relativistic images would mean high accuracy in the determinations of masses and distances of massive deflectors at the centers of galaxies, and 2) the ratio mass to differential time delay for the two outermost relativistic images is almost constant with respect to changes in ($\bt,D_d,D_{ds}$).

In the hope that these numerical solutions would be extended to other massive deflectors, some authors resorted to analytic approximate solutions~\cite{Az,Ait,11a,11,13}. In the case of relativistic images these methods are valid if the angular position $\ta$ of the image is small enough to allow for series expansions. This has been done in~\cite{11a,11,13} for the \S, Reissner-Nordstr\"om, and other \bh and has resulted, among other derivations, in the determination of a log-formula for the angle of deflection and the angular position of the image. Very recently, the method used in~\cite{13} has been applied to gravitational lensing by phantom \BH~\cite{rev}.

In the following we will derive an approximate analytic reference equation for the log-formula for the angle of deflection, which applies to any geometry provided the light motion or a plane projection of it is described by~\eqref{1.2}. We assume that a photon sphere $r_{ps}$ exits. The corresponding \ws radial coordinate is denoted by $\ro_{ps}$, and $\ro_n$ corresponds to $r_n$. Since we are assuming $\ro(r)$ to be a decreasing function of $r$, a scattering case corresponds to $e_3<\ri<e_2=\ro_n$ and $e_2<\ro_{ps}<e_1$. In the strong field limit, as $r_n\to r_{ps}$, both $e_2$ and $e_1$ approach $\ro_{ps}$. If we set\footnote{Had we assumed $\ro(r)$ increasing, a scattering case would correspond to $\ro_n=e_1<\ri$. In the strong field limit we set $\ro_n-\ro_{ps}=\bar{\epsilon}>0$ leading to $\ro_{ps}-e_2=\bar{\epsilon}+O[\bar{\epsilon}]^2$. Eq.~\eqref{s2} would read
\begin{equation*}
    \de\phi=\frac{1}{|\kappa|}\int_{\ro_{ps}+\bar{\epsilon}}^{\ri}\frac{\dd \ro}{\sqrt{(\ro-\ro_{ps}-\bar{\epsilon})(\ro-\ro_{ps}+\bar{\epsilon})(\ro+2\ro_{ps})}}-\pi\,.
\end{equation*}
} $\ro_{ps}-\ro_n=\bar{\epsilon}>0$, then to the relevant first order of approximation we have $e_1-\ro_{ps}=\bar{\epsilon}+O[\bar{\epsilon}]^2$, where $\bar{\epsilon}$ is assumed to be small in order to have relativistic images. With $e_1=\ro_{ps}+\bar{\epsilon}+O[\bar{\epsilon}]^2$, $e_2=\ro_n=\ro_{ps}-\bar{\epsilon}$, and $e_3=-2\ro_{ps}+O[\bar{\epsilon}]^2$ the first equation~\eqref{3.9b} takes the form
\begin{equation}\label{s2}
    \de\phi=\frac{1}{|\kappa|}\int_{\ri}^{\ro_{ps}-\bar{\epsilon}}\frac{\dd \ro}{\sqrt{(\ro_{ps}+\bar{\epsilon}-\ro)(\ro_{ps}-\bar{\epsilon}-\ro)(\ro+2\ro_{ps})}}-\pi\,.
\end{equation}
We introduce the variable $z=(\ro_n-\ro)/(\ro_n-\ri)$. Following~\cite{13} we obtain
\begin{equation}\label{s3}
    \de\phi=-\frac{1}{|\kappa|\sqrt{3\ro_{ps}}}\,\ln\Big[\frac{(\sqrt{3\ro_{ps}}+\sqrt{2\ro_{ps}+\ri})^2}
    {24\ro_{ps}(\ro_{ps}-\ri)}\,
    \bar{\epsilon}\Big]-\pi+\cdots\,.
\end{equation}
The final step is to express $\de\phi$ in terms of $\epsilon$: $r_n-r_{ps}=2M\epsilon$. We have
\begin{equation*}
    \bar{\epsilon}=\ro_{ps}-\ro_n=\ro(r_{ps})-\ro(r_{ps}+2M\epsilon)=
    -2M\ro'(r_{ps})\epsilon+O[\epsilon]^2
\end{equation*}
where $\ro'(r_{ps})=(\dd \ro/\dd r)\big|_{r=r_{ps}}$. Substituting in~\eqref{s3}, we arrive at
\begin{equation}\label{s4}
  \de\phi=-\frac{1}{|\kappa|\sqrt{3\ro_{ps}}}\,\ln\Big[\frac{(\sqrt{3\ro_{ps}}+\sqrt{2\ro_{ps}+\ri})^2}
    {12\ro_{ps}(\ri-\ro_{ps})}\,M\ro'(r_{ps})\epsilon\Big]-\pi+\cdots
\end{equation}
which is valid whether $\ro$ is increasing or decreasing. As we mentioned earlier, this formula applies to all relativistic images of light paths governed by~\eqref{1.2}. The formula looks much easier to handle analytically than that given in~\cite{13}. Using an expansion of $b$ in powers of $\epsilon$, one can determine the positions of the relativistic images if these are too small to allow for series expansions as shown in the following example.

For the \s \abh we have seen that~\cite{GV} $\ro =M/(2r)-1/12$, $\kappa =1$, $\ro_{ps}=1/12$, and $\ri=-1/12$. We obtain $\bar{\epsilon}=M/(2r_{ps})-M/(2r_n)=\epsilon/9+\cdots$, where we have used $r_{ps}=3M$, and finally
\begin{equation*}
    \de\phi=-2\ln\Big[\frac{2+\sqrt{3}}{18}\,\epsilon\Big]-\pi+\cdots
\end{equation*}
which is the same as in~\cite{11a,11}. For small angular positions of the source and its image, the impact parameter $b$ and $\ta$ are related by $b=D_d\ta$. Using an expansion of $b$ in powers of $\epsilon$: $b=3\sqrt{3}M+2\sqrt{3}M\epsilon^2+\cdots$, we arrive at
\begin{equation*}
    \de\phi=-\ln\Big[\frac{D_d\ta}{3\sqrt{3}M}-1\Big]+\ln[216(7-4\sqrt{3})]-\pi+\cdots
\end{equation*}
as in~\cite{13}. The position of the relativistic image of order $n$ in terms of $\bt$ (the angular position of the source) is determined as in~\cite{11a,11,13}.

\setcounter{equation}{0}
\section{Phantom and Normal \RN \BH}\label{5}

It is difficult or impossible to reduce~\eqref{2.13} to~\eqref{1.2} for any $\ga\,$. For $\ga =\pm 2$ and probably for some other values, the problem can be tackled semianalytically in a similar way to what is done in~\cite{Hackmann,Cruz,met}: Limits to the analytical treatment are 1) lack of ``compact" solutions to the polynomial equation (or its polynomial reduced form) $P(r)=0$ and/or 2) lack of solutions to the generally nonpolynomial equation $\De\equiv g_2^3-27g_3^2=0$. All that does not apply to the cases $\ga =1$ and $\ga =0$, which can be entirely analytically solved. In this section we investigate the former case and in the next one we tackle the latter case.

Setting $\ga=1$ in~(\ref{2.6}, \ref{2.9}, \ref{2.13}) with $\epsilon=0$ we obtain
\begin{align}
\label{5.1}&M=\frac{r_++r_-}{2}\,,\; q=\pm \sqrt{\eta_2r_-r_+}\,,\; r_-=\eta_2 |r_-|\,,\; r_-r_+=\eta_2 q^2\\
\label{5.2}&r_{\pm}=M\pm\sqrt{M^2-\eta_2q^2}=M(1\pm\sqrt{1-\eta_2a^2})\\
\label{5.3}&\bigg(\frac{\dd u}{\dd \phi}\bigg)^2=\ell-u^2+2Mu^3-\eta_2q^2u^4
\end{align}
where $\ell=E^2/L^2=1/b^2\geq 0$.

\begin{figure}[h]
\centering
  \includegraphics[width=0.49\textwidth]{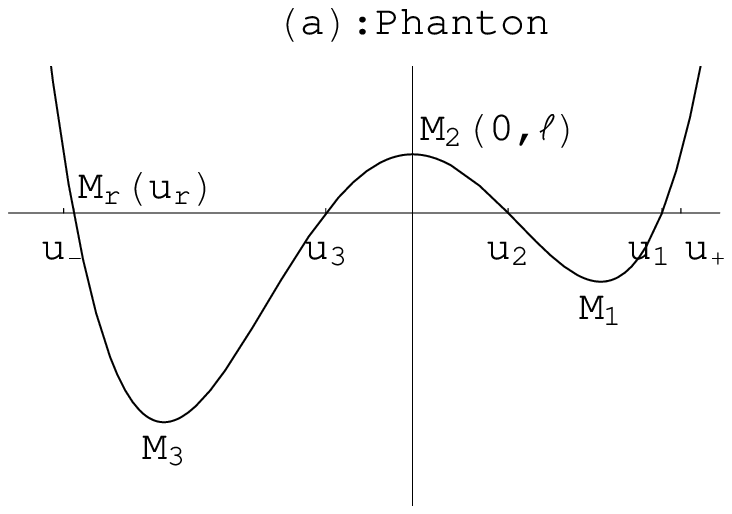}  \includegraphics[width=0.49\textwidth]{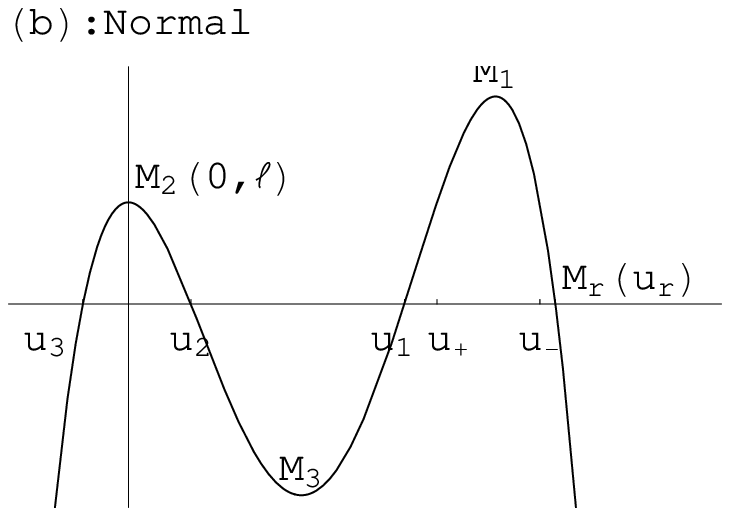}\\
  \caption{\footnotesize{Plots of $Y=\ell-u^2+2Mu^3-\eta_2q^2u^4$ with $\ell \geq 0$. The roots are ($u_r,u_3,u_2,u_1$). $M_r$ is the point with coordinate $(u_r,0)$. (a): The phantom \RN \aBH ($\eta_2=-1$). To perform the reduction of~\eqref{5.3} to~\eqref{1.2} we choose $u_r$ to be the lowest root of $\ell-u^2+2Mu^3+q^2u^4=0$ with $u_-<u_r$.  (b): The normal \RN \aBH ($\eta_2=+1$). To perform the reduction of~\eqref{5.3} to~\eqref{1.2} we choose $u_r$ to be the largest root of $\ell-u^2+2Mu^3-q^2u^4=0$ with $u_-<u_r$. For both plots, ($u_+,u_-$) are the intersections of the graphs of $Y=-u^2+2Mu^3-\eta_2q^2u^4$ with the $u$-axis, which are the same graphs as those shown here but shifted downward $\ell$ units.}}\label{Fig}
\end{figure}

Let $u_r=1/r_r$ be \emph{any} real root of the polynomial in the r.h.s of~\eqref{5.3}:
\begin{equation}\label{5.4}
    \ell-u_r^2+2Mu_r^3-\eta_2q^2u_r^4=0\quad [\ell=\eta_2q^2u_r^2(u_r-u_-)(u_r-u_+)\geq 0]\,.
\end{equation}
Since $\ell\geq 0$, \emph{all} the real roots of the polynomial are either greater than $u_->u_+>0$ or smaller that $u_+$ for $\eta_2=+1$ and they are \emph{all} between $u_-<0$ and $u_+>0$ for $\eta_2=-1$ (Figure~\ref{Fig}).

If $\ell \neq 0$ ($\ell > 0$), we have necessarily $u_r\neq u_+$, $u_r\neq u_-$, and $u_r\neq 0$: It is not possible to find solutions with $\ell \neq 0$ and $r_r =r_+$~\cite[Eqs. (65, 66)]{GV}.

If $u_r$ is a root with multiplicity 1, following the general procedure, we introduce the radial coordinate $y=u-u_r$ followed by $z=1/y$, and finally
\begin{equation}\label{5.5}
    \ro=\frac{3C_1z+C_2}{12}=\frac{C_1}{4(u-u_r)}+\frac{C_2}{12}=-\frac{C_1}{4}\,\frac{r_rr}{r-r_r}+\frac{C_2}{12}\,,\quad \Ta=\phi
\end{equation}
with
\begin{align}
\label{5.5a}&C_1\equiv 2u_r(3Mu_r-1-2\eta_2q^2u_r^2)=2u_r(1-Mu_r)-4\ell/u_r\\
\label{5.5b}&C_2\equiv 6Mu_r-1-6\eta_2q^2u_r^2=5-6Mu_r-6\ell/u_r^2\,.
\end{align}
Eqs.~\eqref{5.5} reduce~\eqref{5.3} to~\eqref{1.2} with
\begin{align}
\label{5.6}&g_2=\frac{1}{12}-\ell Q^2\,,\quad (Q^2=\eta_2 q^2)\\
\label{5.7}&g_3=\frac{1-54\ell M^2+36\ell Q^2}{216}\\
\label{5.8}&\Delta =\ell [M^2(1+36\ell Q^2)-27\ell M^4-Q^2(1+4 \ell Q^2)^2]/16\\
&\;\;\;=-Q^6\ell (\ell-\ell_-)(\ell-\ell_+)=-\eta_2q^6\ell (\ell-\ell_-)(\ell-\ell_+)\nn
\end{align}
where we have used~\eqref{5.4} to eliminate $u_r$ from the expressions of $g_2$, $g_3$. The new parameters ($\ell_-,\ell_+$) are defined by
\begin{equation}\label{5.9}
q^2\ell_{\pm}=\eta _2\,\frac{-27+36 \eta _2 a^2-8 a^4\pm (9-8 \eta _2 a^2)^{3/2}}{32 a^4}\,.
\end{equation}
In the physical case $a^2=q^2/M^2<1$, $0<\ell_+<\ell_-$ for the phantom \abh and $\ell_-<0<\ell_+$ for the normal one.

The transformation~\eqref{5.5} ``splits" the point $r=r_r$ into $r_r^+$ and $r_r^-$ (corresponding to $u_r^-$ and $u_r^+$, respectively). If $C_1<0$, then $r_r^+$ and $r_r^-$ are sent to $\ro=\ro_r=+\infty$ and $\ro=-\infty$, respectively, and if $C_1>0$ the latter limits are reversed. As we shall see in Appendix B, it is always possible to choose the real root $u_r$ so that  $C_1<0$: We choose $u_r$ to be the smallest root for phantom \bh and the largest root for normal ones. The points $\ri$, $\ro_0$ and $\ro_{\pm}$ (corresponding to $r=+\infty$, $r=0$, and $r=r_{\pm}$) on the $\ro$-axis are given by
\begin{align}
\label{5.10}&\ri =\frac{\ell}{2u_r^2}-\frac{1}{12}\,,\;\ro_0=\frac{C_2}{12}=\frac{5}{12}-\frac{\ell}{2u_r^2}-\frac{Mu_r}{2}\\
\label{5.11}&\rho_{\pm}=\frac{C_1}{4(u_{\pm}-u_r)}+\frac{C_2}{12}
\end{align}
which depend on $u_r$ whose analytic expression in terms of ($M,q^2,\ell$) is sizable.

\subsection{Three distinct real roots for $\pmb{w(\ro)=0}$}\label{5.0}

For the phantom case, we derive in Appendix B [Eq.~\eqref{A2}] the following order relations for the $\ro$-parameters:
\begin{equation}\label{B2}
    e_3<\ri<e_2<e_1<\ro_+<\ro_0<\ro_-<\ro_r=+\infty\,.
\end{equation}
The only possible paths are scattering ones from $\ri$ to $e_2$ given by~(\ref{3.9}, \ref{5.5}) or trapped ones from any point $e_1\leq \ro\leq\ro_+$ to the singularity at $\ro_0$ [from any point $r_+\leq r\leq r_1=1/u_1$, where $u_1<u_+$ is the largest root of $\ell-u^2+2Mu^3+q^2u^4=0$ [Figure~\ref{Fig} (a)], to $r=0$]. If the trapped path starts from $\ro=e_1$ its equation is given by~(\ref{3.10}, \ref{5.5}).

For the normal case, Eq.~\eqref{A6} of Appendix B reads
\begin{equation}\label{B6}
    \ro_0<e_3<\ri<e_2<e_1<\ro_+<\ro_-<\ro_r=+\infty\,.
\end{equation}
The only possible paths are scattering ones from $\ri$ to $e_2$ given by~(\ref{3.9}, \ref{5.5}) or many-world ones from any point $e_1\leq \ro\leq\ro_+$ to $\ro=+\infty$ (from any point $r_+\leq r\leq r_1=1/u_1$ to $r_r=1/u_r$ where $u_r>u_-$ is the largest root of $\ell-u^2+2Mu^3-q^2u^4=0$ and $u_1<u_+$ is the second largest root [Figure~\ref{Fig} (b)]). If the many-world path starts from $\ro=e_1$ its equation is given by~(\ref{3.10}, \ref{5.5}).

\subsection{Two distinct real roots for $\pmb{w(\ro)=0}$}\label{5.two}

This corresponds to [Eq.~\eqref{3.12}]
\begin{equation}\label{5.12}
    \frac{1}{12}-\ell \eta_2 q^2>0\;\text{ and } (\ell=0,\,\ell=\ell_-\,\text{ or }\,\ell=\ell_+)\,.
\end{equation}

\subsubsection{Case $\pmb{\ell=0}$.} As we do and explain in Appendix B we choose $u_r=u_-$, which is the smallest (largest) root for the phantom (normal) solution when $\ell=0$. In this case $C_1<0$ and $\ro(r)$ is a decreasing function of $r$ [as $r\to r_-^+$ (from the right),  $\ro\to +\infty$]. The order relations as given in~(\ref{A3}, \ref{A7}) read
\begin{align}
\label{5.14}& \eta_2=-1:\;e_3=e_2=\ri=-\frac{1}{12}<e_1=\ro_+=\frac{1}{6}<\ro_0<\ro_-=+\infty\\
\label{5.13}& \eta_2=+1:\;\ro_0<e_3=e_2=\ri=-\frac{1}{12}<e_1=\ro_+=\frac{1}{6}<\ro_-=+\infty\,.
\end{align}
Since $\ro_0$ is a singularity for the phantom \aBH, there is a trapped path for this hole from $\ro_+\to\ro_0$ given by~(\ref{3.17}, \ref{3.18}) with $\eta_2=-1$.

There is a many-world periodic path for the normal \abh from $\ro_+\to\ro_-$ given by~(\ref{3.16}, \ref{5.5}). Using~\eqref{5.5} with $u=1/r$,  $u_r=u_-$, $C_1=-(r_+-r_-)/r_-^2$, and $C_2=(2r_--3r_+)/r_-$ we have
\begin{equation}\label{5.15}
    4(\ro-e_1)=\frac{r_+-r}{r-r_-}=\frac{(r_+-r)^2}{2Mr-\eta_2 q^2-r^2}\qquad (\eta_2=+1)
\end{equation}
then using the first equation~\eqref{3.16} with $k=\sqrt{e_1-e_3}=1/2$ leads to~(\ref{3.17}, \ref{3.18}).

Had we chosen $u_r=u_+$, instead of $u_r=u_-$, we would reach the same conclusions concerning the nature of the paths. In this case, we would have $C_1=2u_+(1-Mu_+)>0$, and $\ro(r)$ is an increasing function of $r$ [as $r\to r_+^-$ (from the left),  $\ro\to +\infty$] and
\begin{align}
\label{5.17}& \eta_2=-1:\;e_3=e_2=\ri=-\frac{1}{12}<e_1=\ro_-=\frac{1}{6}<\ro_0<\ro_+=+\infty\\
\label{5.16}& \eta_2=+1:\;e_3=e_2=\ri=-\frac{1}{12}<\ro_0<e_1=\ro_-=\frac{1}{6}<\ro_+=+\infty\,.
\end{align}
But instead of~\eqref{5.15} we would obtain
\begin{equation}\label{5.18}
    4(\ro-e_1)=\frac{r-r_-}{r_+-r}=\frac{2Mr-\eta_2 q^2-r^2}{(r_+-r)^2}\;\text{ and }\;\tan\Big[\frac{\phi -C}{2}\Big]=\sqrt{\frac{r-r_-}{r_+-r}}\,.
\end{equation}

\subsubsection{Cases $\pmb{\ell=\ell_{\pm}}$.} Now we consider the cases $\ell=\ell_{\pm}$. Note that in this case $g_2,\;g_3$ are, by~\eqref{5.9}, functions of ($a^2,\eta_2$) only and that $M^2\ell_{\pm}$ are also functions of ($a^2,\eta_2$) only,
\begin{equation}\label{5.18b}
M^2\ell_{\pm}=\eta _2\,\frac{-27+36 \eta _2 a^2-8 a^4\pm (9-8 \eta _2 a^2)^{3/2}}{32 a^6}
\end{equation}
leading to
\begin{equation*}
    \lim_{a^2\to 0}\ell_-=+\infty\;(\eta_2=-1)\;\text{ and } \lim_{a^2\to 0}\ell_+=\frac{1}{27M^2}\;(\eta_2=\pm 1)\,.
\end{equation*}
With $b=\sqrt{1/\ell}$, the last two limits are the \s limit for the impact parameter ($3\sqrt{3}M$) allowing photons to orbit endlessly the hole around the photon sphere without reaching it.

\paragraph{$\pmb{\ell=\ell_+}$.} If $\ell=\ell_+$ and $\eta_2=-1$, $g_2>0$ and $g_3<0$. A similar case has been treated in Eqs.~\eqref{3.19} to~\eqref{3.22}. There is a root with multiplicity 2 at $\ro=e_1=e_2=(1/2)\sqrt{g_2/3}$. The corresponding root $u=u_1$ is such that the r.h.s of~\eqref{5.3} reads: $\ell_+-u^2+2Mu^3+q^2u^4=q^2(u-u_1)^2(u-u_r)(u-u_3)$ where $u=u_3$ corresponds to $\ro=e_3=-2e_1$. This is the case where the point $M_1$ is on the $u$-axis [Figure~\ref{Fig} (a)]. The order relations are given in~\eqref{A4}:
\begin{equation}\label{5.22}
    e_3<\ri<e_1=e_2<\ro_+<\ro_0<\ro_-<\ro_r=+\infty\,.
\end{equation}
There is an unstable circular path at
\begin{equation}\label{5.22b}
    \ro_{ps}=e_1=\frac{\sqrt{(9-8\eta_2 a^2)(9-4\eta_2 a^2-3\sqrt{9-8\eta_2 a^2})}}{24\sqrt{2}a^2}
\end{equation}
(with $\eta_2=-1$) corresponding to $r=r_1=1/u_1$ (the photon sphere) with\footnote{$u_1$ is the largest root of $\ell_+-u^2+2Mu^3+q^2u^4=0$ when $M_1$ is on the $u$-axis (Figure~\ref{Fig} (a)).}
\begin{equation}\label{5.23}
    r_1=r_{ps}=\frac{\sqrt{9-8\eta_2 a^2}+3}{2}\,M>3M>r_+
\end{equation}
and  spiral paths from $r=+\infty$ ($\ro=\ri$) to $r=r_1$ ($\ro=e_1$) and from $r=r_+$ ($\ro=\ro_+$) to $r=r_1$ ($\ro=e_1$) given by~\eqref{3.22} and~\eqref{5.5} with $k=\sqrt{3e_1}$. There is also a trapped path from $r=r_1$ ($\ro=e_1$) to $r=0$ ($\ro=\ro_0$) given by~\eqref{3.22} with the $+$ sign. In the limit $a^2\to 0$, $r_1\to 3M$, which is the \s limit.

If $\ell=\ell_+$ and $\eta_2=+1$, $g_2>0$ and $g_3<0$. There is a root with multiplicity 2 at $\ro=e_1=e_2=(1/2)\sqrt{g_2/3}$. The corresponding root $u=u_1$ is such that the r.h.s of~\eqref{5.3} reads $\ell_+-u^2+2Mu^3-q^2u^4=q^2(u-u_1)^2(u-u_r)(u-u_3)$, where $u=u_3$ corresponds to $\ro=e_3=-2e_1$. This is the case where the point $M_3$ is on the $u$-axis [Figure~\ref{Fig} (b)]. The order relations are given in~\eqref{A8}:
\begin{equation}\label{5.25}
\ro_0<e_3<\ri<e_1=e_2<\ro_+<\ro_-<\ro_r=+\infty\,.
\end{equation}
There is an unstable circular path at $\ro=e_1$, given by~\eqref{5.22b} taking $\eta_2=+1$, which corresponds to $r_{ps}=r_1=1/u_1$ (the photon sphere)\footnote{$u_1$ is the smallest positive root of $\ell_+-u^2+2Mu^3-q^2u^4=0$ when the point $M_3$ is on the $u$-axis (Figure~\ref{Fig} (b)).}. The latter is given by~\eqref{5.23} taking $\eta_2=+1$, leading to $r_+<r_{ps}<3M$. There are  spiral paths from $r=+\infty$ ($\ro=\ri$) to $r=r_1$ ($\ro=e_1$) and from $r=r_+$ ($\ro=\ro_+$) to $r=r_1$ ($\ro=e_1$) given by~\eqref{3.22} and~\eqref{5.5} with $k=\sqrt{3e_1}$. There is also a many-world periodic bound path from $r>r_+$ through $r=r_-$ to $r=r_r>0$, which emerges in another copy of the space-time after crossing $r=r_-$. This is also given by~\eqref{3.22} with the $+$ sign. In the limit $a^2\to 0$, $r_1\to 3M$, which is the \s limit.

\paragraph{$\pmb{\ell=\ell_-}$.} We have necessarily $\eta_2=-1$ since $\ell_-<0$ for the normal \RN \aBH. In this case, the r.h.s of~\eqref{5.3}, $\ell_--u^2+2Mu^3+q^2u^4$, has only one real root\footnote{$u_1$ is the only real root of $\ell_--u^2+2Mu^3+q^2u^4=0$ when the point $M_3$ is on the $u$-axis (Figure~\ref{Fig} (a)).} $u_1=-[\sqrt{9+8a^2}+3]/(4a^2M)<0$ with multiplicity 2 and two complex roots\footnote{In this case the reduction of~\eqref{5.3} does not lead to~\eqref{1.2}; rather, it leads to a similar equation with an irreducible quadratic form on the r.h.s., a polynomial of degree 2 with complex roots.}. Thus, the r.h.s of~\eqref{5.3} is always positive with only absorbed paths from spatial infinity to the singularity at $r=0$ given by~\eqref{3.11} and~\eqref{5.5}.

\subsection{One real root for $\pmb{w(\ro)=0}$}\label{5.1}

This corresponds to [Eq.~\eqref{3.23}]
\begin{align}
\label{5.27}&\frac{27+36 a^2+8 a^4- (9+8  a^2)^{3/2}}{32 M^2a^6}<\ell<\frac{27+36  a^2+8 a^4+ (9+8 a^2)^{3/2}}{32 M^2a^6}\qquad (\eta_2=-1)\\
\label{5.28}&\ell>\frac{36 a^2-27-8 a^4+ (9-8 a^2)^{3/2}}{32 M^2a^6}\qquad (\eta_2=+1)\,.
\end{align}

For the phantom solution ($\eta_2=-1$), this is the case where the point $M_1$ is above the $u$-axis and $M_3$ is below it [Figure~\ref{Fig} (a)]. The two real roots ($u_r<u_3$) of $\ell-u^2+2Mu^3+q^2u^4=0$ are negative and ($u_1,u_2$) are now complex roots, so ($e_1,e_2$) no longer exist . Eqs.~(\ref{A1}, \ref{A2}) become
\begin{align}
&u_-<u_r<u_3<0<u_+\nn\\
\label{5.29}&e_3<\ri<\ro_+<\ro_0<\ro_-<\ro_r=+\infty
\end{align}
with only absorbed paths from spatial infinity to the singularity at $r=0$ given by~\eqref{3.11} and~\eqref{5.5}.

For the normal solution ($\eta_2=+1$), this is the case where the point $M_3$ is above the $u$-axis [Figure~\ref{Fig} (b)]. The two real roots of $\ell-u^2+2Mu^3-q^2u^4=0$ satisfy $u_3<0$ and $u_r>u_->0$ is the largest one. ($u_1,u_2$) are now complex roots, so ($e_1,e_2$) no longer exist.  Eqs.~(\ref{A5}, \ref{A6}) become
\begin{align}
& u_3<0<u_+<u_-<u_r\nn\\
\label{5.30}&\ro_0< e_3<\ri<\ro_+<\ro_-<\ro_r=+\infty\,.
\end{align}
The only existing paths are two-world scattering paths from spatial infinity to $r=r_r=1/u_r>0$ given by~\eqref{3.11} and~\eqref{5.5}.

The log-formula for the deflection angle is easily determined using~\eqref{s4} with $\ri$, $M^2\ell_+$, and $\ro_{ps}$ given by~(\ref{5.10}, \ref{5.18b}, \ref{5.22b}), respectively, $\kappa = 1$, and $U_r=Mu_r$ is the lowest root (if $\eta_2=-1$) or largest one (if $\eta_2=+1$) of $M^2\ell_+-U^2+2U^3-\eta_2a^2U^4=0$.

\setcounter{equation}{0}
\section{Null geodesics of phantom and normal EMD}\label{4}

In this section we restrict ourselves to the case $\ga =0$, which corresponds to $\eta_1=+1$, and then~\eqref{2.3} implies $\eta_2=-1$ for the cosh solution and  $\eta_2=+1$ for the sinh one. Thus we will be considering E-anti-MD for the cosh solution and normal EMD for the sinh one.

Instead of $u=(1/r_-)-(f_-/r_-)$, we use $f_-$ as a radial coordinate. This way we reduce~\eqref{2.13} for light paths ($\epsilon =0$) to
\begin{equation}\label{4.1}
  \bigg(\frac{\dd f_-}{\dd \phi}\bigg)^2=[\al f_-^3 - (3\al +1)f_-^2 + (3\al + \bt +2)f_- - (\al +1)]f_-
\end{equation}
where, using~\eqref{2.8},
\begin{equation}\label{4.2}
    \al \equiv -\frac{r_+}{r_-}=-\eta_2\,\frac{2M^2}{q^2}=-\eta_2\,\frac{2}{a^2}\,,\;\;\bt \equiv \frac{r_-^2E^2}{L^2}=\frac{q^4E^2}{M^2L^2}=\frac{q^4}{M^2b^2}\geq 0\,.
\end{equation}
In the physical case $a^2<1$, to which we restrict ourselves, $\al$ is constrained by
\begin{equation}\label{4.3}
    \al >2 \,\text{ if}\, \eta_2 =-1\,,\;\;\al <-2 \,\text{ if}\, \eta_2 =+1
\end{equation}
for the phantom cosh and normal sinh solutions, respectively.

The next step is to introduce the variable $R=1/(f_--f_0)$ where $f_0$ is a zero of the fourth order polynomial in $f_-$ on the r.h.s of~\eqref{4.1}. We choose $f_0=0$, leading to
\begin{equation*}
    \bigg(\frac{\dd R}{\dd \phi}\bigg)^2=\al - (3\al +1)R + (3\al + \bt +2)R^2 - (\al +1)R^3\,.
\end{equation*}
The final steps consist in eliminating the term in $R^2$ and rescaling $\phi$ by introducing the \ws coordinates ($\ro,\Ta$) defined by
\begin{align}
\label{4.4}& R=-\frac{4^{1/3}}{(\al +1)^{1/3}}\,\ro + \frac{3\al + \bt +2}{3(\al +1)}\\
\label{4.5}& \dd \Ta = -\eta_2\,\frac{(\al +1)^{1/3}}{4^{1/3}}\,\dd \phi\,,\qquad (\dd \phi \cdot \dd \Ta >0)
\end{align}
and $\dd \phi \cdot \dd \Ta >0$ by~\eqref{4.3}. The reduced equation is~\eqref{1.2}: $(\dd \ro/\dd \Ta)^2=4\ro^3-g_2\ro-g_3$ with
\begin{align}
\label{4.6}&g_2=\frac{4^{1/3}}{3} \frac{1+2(2+3\alpha )\beta +\beta ^2}{(1+\alpha )^{4/3}}\\
\label{4.7}&g_3=\frac{2-3(5+12\alpha +9\alpha ^2)\beta -6(2+3\alpha )\beta ^2-2\beta ^3}{27(1+\alpha )^2}\,.
\end{align}

Note that, if $\al$ is restricted by~\eqref{4.3}, $\ro(r)$ is a decreasing function of $r$ for all $\eta_2$. $\ro(r)$ and its inverse function are given by
\begin{equation}\label{4.8}
    \rho =\frac{(\beta -1)r-(3\alpha +\beta +2)r_-}{3\cdot 4^{1/3}(1+\alpha )^{2/3}(r-r_-)}\;\text{ and }\;r=\frac{r_-[3\alpha +\beta +2-3\cdot 4^{1/3}(1+\alpha )^{2/3}\rho ]}{\beta -1-3\cdot 4^{1/3}(1+\alpha )^{2/3}\rho }
\end{equation}
so that, using $r_-=\eta_2 |r_-|$ and $\al+1=-\eta_2|\al+1|$, we arrive at
$\dd \ro/\dd r=-|r_-||\al+1|^{1/3}/[4^{1/3}(r-r_-)^2]$.

In the limit $r\to r_-$, $\ro\to -3r_-(\al+1)/(r-r_-)=3|r_-||\al+1|/(r-r_-)$ for all $\eta_2$. Thus the transformation~\eqref{4.8} ``splits" the point $r_-$ into $r_-^-$ and $r_-^+$ and sends the point $r_-^-$ to $\ro=-\infty$ and the  point $r_-^+$ to $\ro=\ro_-=+\infty$. The points $\ri$, $\ro_0$, and $\ro_+$ (corresponding to $r=+\infty$, $r=0$, and $r=r_+$) on the $\ro$-axis are given by
\begin{equation}\label{4.9}
    \ri =\frac{\bt -1}{3\cdot 4^{1/3}(\al+1)^{2/3}}\,,\;\ro_0=\frac{3\al+\bt+2}{3\cdot 4^{1/3}(\al+1)^{2/3}}\,,\;
    \rho _+=\frac{\beta +2}{3\cdot 4^{1/3}(1+\alpha )^{2/3}}
\end{equation}
and $\ro_0>0$ for phantom \BH. If ($e_1,e_2,e_3$) are real, the order relations of these roots with respect to ($\ri,\ro_0,\ro_+$) depend on $(\al,\bt)=\vec{p}$. This will be done for each case (phantom or normal) separately.

Ordering ($\ri,\ro_0,\ro_+$) is also done separately as follows. For the phantom cosh \abh we have $r_-^+<0<r_+<+\infty$, which leads to [$\ro(r)$ is always decreasing]
\begin{equation}\label{4.9a}
\ri<\ro_+<\ro_0<\ro_-=+\infty\,.
\end{equation}
For the normal sinh \abh we have $0<r_-^-<r_-^+<r_+<+\infty$. But since $\ro(r)$ is always decreasing, if one moves on the $r$-axis along the path $r=+\infty$ $\to$ $r_+$ $\to$ $r_-^+$ $\to$ $r_-^-$ $\to$ $0$, the corresponding point on the $\ro$-axis moves along the path $\ri$ $\to$ $\ro_+$ $\to$ $\ro_-=+\infty$ $\to$ (in a circular rotation) $-\infty$ $\to$ $\ro_0$. Thus
\begin{equation}\label{4.9b}
\ro_0<\ri<\ro_+<\ro_-=+\infty\,.
\end{equation}

Let ($\bt_1,\bt_2,\bt_3,\bt_4$) be the following $\al$ functions:
\begin{align}
\label{4.10}& \bt_{1,\,2}=-(2+3\al)\mp \sqrt{3(1+\al)(1+3\al)}\qquad (1\to -,\;2\to +)\\
\label{4.11}& \bt_{3,\,4}=\frac{1-18\al-27\al^2\pm (1+9\al)\sqrt{(1+\al)(1+9\al)}}{8\al}\qquad (3\to +,\;4\to -)
\end{align}
in terms of which we have
\begin{align}
\label{4.12}&g_2=\frac{4^{1/3}}{3} \frac{(\bt-\bt_1)(\bt-\bt_2)}{(1+\alpha )^{4/3}}\\
&\De\equiv g_2^3-27g_3^2=\frac{\bt [4+(1-18\al-27\al^2)\bt -4\al\bt^2]}{(1+\al)^2}=\frac{-4\al\bt(\bt-\bt_3)(\bt-\bt_4)}{(1+\al)^2}\nn\,.
\end{align}

\subsection{The phantom cosh \aBH: $\pmb{\al>2,\;\eta_2=-1}$}

In this case $g_2>0$ for all $\bt\geq 0$ (Eq.~\eqref{4.2}), $\bt_4<0$ and $\bt_3>0$.

\subsubsection{Three distinct real roots for $\pmb{w(\ro)=0}$}

This corresponds to (Eq.~\eqref{3.3}): $0<\bt<\bt_3$ which leads, using~(\ref{3.4}, \ref{4.9}, \ref{4.9a}), to
\begin{equation}\label{4.14}
    e_3<\ri<e_2<e_1<\ro_+<\ro_0<\ro_-=+\infty \quad (\ri<0)\,.
\end{equation}
To order ($\ri,\ro_0,\ro_+$)  with respect to  ($e_1,e_2,e_3$), as done in~\eqref{4.14}, we may use different methods as plotting the surfaces $\ro_+-e_1$ and so on or simply evaluate the \ws polynomial and its derivatives $w'=12\ro^2-g_2$ and $w''=24\ro$ at ($\ri,\ro_0,\ro_+$). For instance, $w(\ro_+)>0$, $w'(\ro_+)>0$ and $w''(\ro_+)>0$.

This case has been treated in Subsection~\ref{3.1.1} case (b). Since $\ro_0$ is a singularity for the cosh \aBH, there is a trapped path from $e_1$ to $\ro_0$ given by~(\ref{3.10}, \ref{4.5}, \ref{4.8}). The scattering path from $\ri\to e_2\to\ri$ is given by~(\ref{3.9}, \ref{4.5}, \ref{4.8}).

\subsubsection{Two distinct real roots for $\pmb{w(\ro)=0}$}

This corresponds to [Eq.~\eqref{3.12}]
\begin{equation}\label{4.15}
    \bt=0\;\text{ or }\;\bt=\bt_3\,.
\end{equation}

For $\bt=0$, $\ro_+=e_1$ and $g_3>0$ so that $g_3=(g_2/3)\sqrt{g_2/3}$. The relations~\eqref{4.14} are still valid, in the limit we have $e_3=\ri=e_2$. This case has been treated in Eqs.~\eqref{3.13} to~\eqref{3.16}. Since $\ro_0$ is a singularity for the cosh \aBH, there is a trapped or terminating bound path from $\ro_+=e_1$ ($r=r_+$) to $\ro_0$ ($r=0$) given by~(\ref{3.16}, \ref{4.5}, \ref{4.8}) with $e_1=-2e_3=-2e_2=-2\ri=2^{1/3}/[3(1+\al)^{2/3}]$:
\begin{equation*}
    2^{2/3}(1+\al)^{2/3}\ro=\frac{2}{3}+\tan^2\Big[\frac{\Ta-C}{2^{1/3}(1+\al)^{1/3}}\Big]\,.
\end{equation*}
Substituting $\bt=0$ into~\eqref{4.6} and then into~\eqref{3.14} and the second equation~\eqref{4.8}, we obtain the radius of the stable circular path at $r=\infty$, as in the \s case.

For $\bt=\bt_3$, $g_3<0$ so that $g_3=-(g_2/3)\sqrt{g_2/3}$. This case has been treated in Eqs.~\eqref{3.19} to~\eqref{3.22}. The relations~\eqref{4.14} remain valid with $2e_1=2e_2=-e_3=\sqrt{g_2(\bt_3)/3}$. There are spiral paths given by~(\ref{3.22}, \ref{4.5}, \ref{4.8}) which approach the unstable circular path at $\ro=e_1$ from above($\ri$)/below($\ro_0$). The radii $\ro_{ps}=e_1$ and $r_{ps}=r_1$ of the unstable circular path (photon sphere) are given by
\begin{align}
\label{4.15b}& \ro_{ps}=\frac{1}{24}\Big\{\frac{(1+9\al)[1+9\al^2-\eta_2\sqrt{A}+\al(2+3\eta_2\sqrt{A})]}
{2^{1/3}\al^2(1+\al)^{1/3}}\Big\}^{1/2}\\
\label{4.16}& r_{ps}=\frac{8r_+(1+3\alpha )}{1-\eta_2\sqrt{A}-\alpha (10+27\alpha +9\eta_2\sqrt{A})-\eta_2\sqrt{2A}\sqrt{1-\eta_2\sqrt{A}+\alpha (2+9\alpha +3\eta_2\sqrt{A})}}
\end{align}
(with $\eta_2=-1$ and $\al>2$) where $A=(1+\alpha )(1+9 \alpha )$ and $r_+=2M$. The limit $q^2\to 0$ corresponds to $\al\to +\infty$. The radius $r_{ps}$, as given by~\eqref{4.16}, decreases from $(5+\sqrt{57})r_+/8$ to the \s limit $3r_+/2$ as $\al$ increases from $2\to +\infty$.

\subsubsection{One real root for $\pmb{w(\ro)=0}$}

This case corresponds to [Eq.~(\ref{3.23})]: $\bt>\bt_3$ leading to
\begin{equation}\label{4.23b}
    \ri<e_r<\ro_+<\ro_0<\ro_-=+\infty
\end{equation}
where $e_r$ is the real given by~\eqref{4.23c}. For $\al>2$ it is not possible to have $g_2=0$ and $g_3=0$, so there is no solution $e_r=0$ with multiplicity 3.
Since $\ro_0$ is a singularity for the cosh \aBH, there is a trapped path from $e_r$ to $\ro_0$ for the generic case $\bt>\bt_3$ given by~(\ref{3.10}, \ref{4.5}, \ref{4.8}).

\subsection{The normal sinh \aBH: $\pmb{\al<-2,\;\eta_2=+1}$}

In this case $0<\bt_4<\bt_1<\bt_2<\bt_3$. Thus, the condition $\De>0$ [Eq.~\eqref{4.12}] ensures $g_2> 0$.

\subsubsection{Three distinct real roots for $\pmb{w(\ro)=0}$}

This corresponds to [Eq.~\eqref{3.3}]: $0<\bt<\bt_4\;\text{ or }\;\bt>\bt_3$, which leads, using~(\ref{3.4}, \ref{4.9}, \ref{4.9b}), to
\begin{align}
\label{4.18}&\ro_0<e_3<\ri<e_2<e_1<\ro_+<\ro_-=+\infty\;\text{ if }\;0<\bt<\bt_4\quad (\ri<0)\\
\label{4.19}&e_3<e_2<\ro_0<e_1<\ri<\ro_+<\ro_-=+\infty\;\text{ if }\;\bt>\bt_3\,.
\end{align}

For the case $0<\bt<\bt_4$, which has been treated in Section~\ref{3.1.1} case (d), the solution for the scattering path from $\ri\to e_2\to\ri$ ($r=\infty\to r_2\to r=\infty$) is given by~(\ref{3.9}, \ref{4.5}, \ref{4.8}). There is another path from $r=r_1$ ($\ro=e_1$) to $r=r_-$ ($\ro=\ro_-$), which is a trapped path where $\ro_-$ is a null singularity for the sinh \aBH. If we choose $\Ta =0$ at $\ro=e_1$, then the solution is given by~(\ref{3.10}, \ref{4.5}, \ref{4.8}).

The case $\bt>\bt_3$ has been treated in Section~\ref{3.1.1} case (c). Since $\ro_-$ is a singularity, we have an absorbed path from spatial infinity to the singularity. The solution is given by~(\ref{3.11}, \ref{4.5}, \ref{4.8}).

\subsubsection{Two distinct real roots for $\pmb{w(\ro)=0}$}

This corresponds to [Eq.~\eqref{3.12}]: $\bt=0\,,\;\bt=\bt_4\;\text{ or }\;\bt=\bt_3$.

The discussion of the case $\bt=0$ for the normal sinh \abh is similar to that for the phantom cosh one. The information in the first paragraph following~\eqref{4.15} applies to this case if we replace ``$\ro_0$" by ``$\ro_-$",  ``$r=0$" by ``$r=r_-$" and ``cosh" by ``sinh". Thus, there is a trapped path from $\ro_+$ to the singularity $\ro_-$ given by~(\ref{3.16}, \ref{4.5}, \ref{4.8}).

For $\bt=\bt_4$, $g_3<0$ so that $g_3=-(g_2/3)\sqrt{g_2/3}$. This case corresponds to the case $\bt=\bt_3$ of the phantom cosh \aBH; the discussion in the second paragraph following~\eqref{4.15} applies and the radii of the unstable circular path (photon sphere) are obtained from~(\ref{4.15b}, \ref{4.16}) taking $\eta_2=+1$ ($\al<-2$). The limit $q^2\to 0$ corresponds to $\al\to -\infty$. The radius $r_{ps}$ increases from $(7+\sqrt{17})r_+/8>r_+$ to the \s limit $3r_+/2$ as $|\al|$ increases from $2\to +\infty$ ($\al$ decreases from $-2\to -\infty$).

For $\bt=\bt_3$, $g_3>0$ so that $g_3=+(g_2/3)\sqrt{g_2/3}$. In this case $e_3=e_2=-2e_1$ [Eq.~\eqref{3.13}], and all remaining inequalities in~\eqref{4.19} are still valid. Since $\ro_-$ is a singularity, we have an absorbed path from spatial infinity to the singularity. The solution is given by~(\ref{3.11}, \ref{4.5}, \ref{4.8}). The value $r_3=r_2$ corresponding to $e_3=e_2$, which should give the radius of the stable circular path, is such that $0<r_3=r_2<r_-=r_{\text{sing}}$.

\subsubsection{One real root for $\pmb{w(\ro)=0}$}

This case corresponds to [Eq.~(\ref{3.23})]: $\bt_4<\bt<\bt_3$ leading to
\begin{equation}\label{4.23}
    \ro_0<e_r<\ri<\ro_+<\ro_-=+\infty
\end{equation}
where $e_r$ is the real root given by~\eqref{4.23c}. For $\al<-2$ it is not possible to have $g_2=0$ and $g_3=0$, so there is no solution $e_r=0$ with multiplicity 3. In the generic case $\bt_1\leq\bt\leq\bt_2$ and $g_3\neq 0$ there is an absorbed path from $\ri$ to the singularity at $\ro=\ro_-=+\infty$ given by~(\ref{3.11}, \ref{4.5}, \ref{4.8}).

The log-formula for the deflection angle is easily determined using~\eqref{s4} with $\ri$ and $\ro_{ps}$ given by~(\ref{4.9}, \ref{4.15b}), respectively, and $|\kappa| = |\al +1|^{1/3}/4^{1/3}$.

\section{Conclusion}

To the first order of approximation, all \bh of phantom and normal EMD deflect light in the same manner. If we restrict ourselves to physical conditions [$a^2\leq 1$ for $\eta_2=-1$ and  $a^2\leq (1+\ga)/(2\ga)$ for $\eta_2=+1$], then 1) for $\ga$ larger than some $\ga_0$, which is likely in ($-0.2,0$) and depending on the parameters of the \aBH, \bh of E-anti-M-(anti)-D theory (regardless of the sign of $\eta_1$) cause light rays to deflect with larger angles than \bh of EM-(anti)-D. The difference in the angles and the relative discrepancy ever increase with $a^2$ for fixed ($u_n,\ga$). For fixed ($a^2,\ga$), they increase with $1/r_n$ and diverge as $r_n$ approaches the photon sphere of E-anti-M-(anti)-D \BH. 2) For $\ga<\ga_0$ and $\eta_2\ga R_{\de\phi}<0$ light is more deflected by the \bh of EMD than by those of E-anti-MD; the relative discrepancy for larger values of the impact parameter is, however, much larger for the \bh of E-anti-MD.

Time delay and relativistic images are other ingredients, besides deflection, allowing for the determination of the nature of matter. From this point of view a very useful log-formula for the positions of the images has been determined.

The method based on the \ws polynomial to integrate geodesic motion and determine exact solutions is equivalent to other methods using potential barriers and can be applied systematically. The advantage of using the method based on the \ws polynomial is that motion is allowed in at most two regions: In between the smallest root of the polynomial and the intermediate one and/or for values greater than the largest root. This highly simplifies the problem. Some of the systematic resolutions consist as follows: 1) The angle of deflection and the log-formula have standard expressions for all problems that can be brought to \ws differential equation. 2) If the smallest and intermediate roots of the \ws polynomial are equal for some value of the vector of parameters, there should be a stable circular path for the corresponding radial coordinate $r$ if the latter is within accessible limits to observers. 3) If the largest and intermediate roots are equal for some value of the vector of parameters, there should be an unstable circular path (photon sphere) for the corresponding radial coordinate $r$ if the latter is within accessible limits to observers. 4) Existence of spiral paths, which approach endlessly the photon spheres, is a consequence of 3). 5) Existence and identification of divergencies for the angle of deflection: a logarithmic one if 3) holds or a power law one (to the power $-1/2$) if the three real roots are zero. 6) Ordering of the parameters expressing spatial infinity, singularity, horizons and so on on the \ws axis is derived by circular rotation (from their given order relations on the $r$-axis) in the one or the other way depending on the coordinate transformation relating the \ws radial coordinate to the spherical radial one (increasing or decreasing).

Phantom \RN \bh are characterized by the existence of trapped and absorbed null paths that do not exist for normal \RN \BH. Their other noncommon paths include many-world (periodic bound) and two-world paths that exist only for normal \RN \BH. Their common paths include scattering, spiral (existence of logarithmic divergencies), and unstable circular paths with radii approaching the \s limit from above for phantom \bh and from below for normal ones.

Both phantom cosh and normal sinh \bh of EMD theory are characterized by the presence of scattering, trapped, and unstable circular paths, thus spiral paths and the existence of logarithmic divergencies. The photon spheres are larger or smaller than the \s one, respectively, and approach it in the limit of no electric charge. The phantom solution has no absorbed path while the normal one does.

\section*{Acknowledgments}

Thanks are due to G\'erard Cl\'ement (LAPTH) for helpful correspondence.

\section*{ Appendix A: Geodesic equations and angle of deflection}
\renewcommand{\theequation}{A.\arabic{equation}}
\setcounter{equation}{0}

Related to the two Killing vectors ($\partial_t,\partial_{\phi}$) are the two constants of motion ($E,L$) given by
\begin{equation}\label{2.10}
f_{+}f_{-}^{\gamma}\,\frac{\dd t}{\dd \tau} = E\,,\;r^2\sin^2\ta f_{-}^{1-\gamma}\,\frac{\dd \phi}{\dd \tau} = L\,.
\end{equation}
Since~\eqref{2.2} is endowed with spherical symmetry, the motion happens in a plane through the origin. Letting the plane be $\ta=\pi/2$ and inserting~(\ref{2.10}) into the line element~\eqref{2.2}, we bring it to (with $\epsilon=1,\,0$ for massive, massless particles, respectively)
\begin{equation}\label{2.12}
  \bigg(\frac{\dd r}{\dd \tau}\bigg)^2=E^2-f_{+}f_{-}^{\gamma}\bigg[\epsilon+\frac{L^2}{r^{2}f_{-}^{1-\gamma}}\bigg].
\end{equation}
For scattering states $E^2-\eps>0$. Eliminating $\tau$ in~(\ref{2.10}, \ref{2.12}) and using $u=1/r$ we arrive at
\begin{equation}\label{2.13}
  \bigg(\frac{\dd u}{\dd \phi}\bigg)^2=\frac{f_{-}^{2(1-\gamma)}}{L^2}\Bigg[E^2-f_{+}f_{-}^{\gamma}
  \bigg[\epsilon+\frac{L^2u^2}{f_{-}^{1-\gamma}}\bigg]\Bigg].
\end{equation}

From now on we take $\eps=0$ so that the condition for light scattering is $E^2>0$. Now, let $g(u):=u^2f_+f_{-}^{2\gamma-1}$ and $u_n=1/r_n$ be the point on the light scattering geodesic nearest the origin where $\frac{\dd u}{\dd \phi}(u_n)=0$. Since $E^2=L^2g(u_n)$, the angle of deflection, which is twice the variation of $\phi$ minus $\pi$, takes the form
\begin{equation}\label{2.14}
    \delta\phi = 2\int_{0}^{u_n}\frac{\dd u}{f_{-}^{1-\gamma}(u)\sqrt{g(u_n)-g(u)}}-\pi =2\int_0^1\frac{u_n\sqrt{1-x^2}}{f_{-}^{1-\gamma}(u_nx)\sqrt{g(u_n)-g(u_nx)}}\,\frac{\dd x}{\sqrt{1-x^2}}-\pi
\end{equation}
where $u=u_nx$. If $u_n\ll 1$, corresponding to scattering with large values of the impact parameter ($b=L/E$),
\begin{multline*}
\frac{u_n\sqrt{1-x^2}}{f_{-}^{1-\gamma}(u_nx)\sqrt{g(u_n)-g(u_nx)}}=1+
\bigg[\frac{[(\ga-1)r_-+2M]}{2}\,\frac{1-x^3}{1-x^2}-(\ga-1)r_-x\bigg]u_n\\
+\frac{1}{8(1+x)^2}\bigg\{3r_+^2(1+x+x^2)^2+2r_-r_+ [2\gamma -1+2\gamma x+(6\gamma -1)x^2+2 x^3+x^4]\\
+r_-^2 [4\gamma ^2-1+6(2\gamma
-1)x+(8\gamma ^2+1)x^2+6 x^3+3 x^4]\bigg\}\,u_n^2+O[u_n]^3
\end{multline*}
where we have used~\eqref{2.6}: $r_++\ga r_-=2M$. Performing the integrations over $x$ we obtain~\eqref{2.16}.

\section*{ Appendix B: Order relations for the phantom and normal \RN \BH}
\renewcommand{\theequation}{B.\arabic{equation}}
\setcounter{equation}{0}

\subsection*{The phantom case $\pmb{\eta_2=-1}$}

In the case where all four roots of $\ell-u^2+2Mu^3+2q^2u^4=0$ have multiplicity 1 we can choose any root to perform the reduction of~\eqref{5.3} to~\eqref{1.2}. For the phantom \RN \abh we choose $u_r$ to be the smallest root as shown in Figure~\ref{Fig} (a)
\begin{equation}\label{A1}
    u_-<u_r^-<u_r^+<u_3<0<u_2<u_1<u_+<+\infty\,.
\end{equation}
As defined in the first expressions of Eqs.~(\ref{5.5a}, \ref{5.5b}), ($C_1,C_2$) are proportional to the first and second derivatives of $\ell-u^2+2Mu^3+q^2u^4$ at $u=u_r$, respectively. At the point $M_r(u_r,0)$ of Figure~\ref{Fig} (a), the function is decreasing and concave up (convex), thus $C_1<0$, $C_2>0$, and $\ro$ is an increasing function of $u$ (a decreasing function of $r$).

Since $C_1<0$, the coordinate transformation~\eqref{5.5} ``splits" the point $u_r$ into $u_r^+$ and $u_r^-$, which correspond to $\ro=-\infty$ and $\ro=+\infty$, respectively. If one starts to move on the $u$-axis from the right to the left: from $u=+\infty$ ($r=0$) to $u_+$ to $u_1$ to $\cdots$ to $u_r^+$ to $u_r^-$ and finally to $u_-$. Since $\ro$ is an increasing function of $u$, the corresponding point on the $\ro$-axis starts to move from $\ro_0$ to $\ro_+$ to $e_1$ to $\cdots$ to $\ro=-\infty$ and then back in a circular rotation to $\ro=+\infty$ and finally to $\ro_-$. Thus, we have the following order relations for the $\ro$-parameters:
\begin{equation}\label{A2}
   -\infty< e_3<\ri<e_2<e_1<\ro_+<\ro_0<\ro_-<\ro_r=+\infty\,.
\end{equation}
(Of course, this ordering can be derived by algebraic methods).

If $\ell=0$ ($M_2$ on the $u$-axis), then $u_r=u_-$, $u_2=u_3=0$, and $u_1=u_+$ ($\ro_r=+\infty$, $e_2=e_3$, and $e_1=\ro_+$). Using~(\ref{3.13}, \ref{5.5a}, \ref{5.5b}, \ref{5.6}, \ref{5.7}, \ref{5.10}, \ref{5.11}) we obtain $12\ro_0=5-[6\eta_2(1+\sqrt{1-\eta_2a^2})/a^2]$ (with $\eta_2=-1$), and thus
\begin{equation}\label{A3}
   e_3=e_2=\ri=-\frac{1}{12}<e_1=\ro_+=\frac{1}{6}<\ro_0<\ro_-=+\infty\,.
\end{equation}

If $\ell=\ell_+$, then $u_1=u_2$ ($M_1$ on the $u$-axis) and~\eqref{A2} becomes
\begin{equation}\label{A4}
   -\infty< e_3<\ri<e_2=e_1<\ro_+<\ro_0<\ro_-<\ro_r=+\infty\,.
\end{equation}

If $\ell=\ell_-$, then $u_r=u_3$ ($M_3$ on the $u$-axis) and the root $u_r$ has multiplicity 2.

\subsection*{The normal case $\pmb{\eta_2=+1}$}

In the case where all four roots of $\ell-u^2+2Mu^3-2q^2u^4=0$ have multiplicity 1 we can choose any root to perform the reduction of~\eqref{5.3} to~\eqref{1.2}. For the normal \RN \abh we choose $u_r$ to be the largest root as shown in Figure~\ref{Fig} (b)
\begin{equation}\label{A5}
    u_3<0<u_2<u_1<u_+<u_-<u_r^-<u_r^+<+\infty\,.
\end{equation}
At the point $M_r(u_r,0)$ of the (b) plot the function is decreasing and concave down (concave), and thus $C_1<0$, $C_2<0$, and $\ro$ is an increasing function of $u$ (a decreasing function of $r$). Since $C_1<0$, the coordinate transformation~\eqref{5.5} splits the point $u_r$ into $u_r^+$ and $u_r^-$, which correspond to $\ro=-\infty$ and $\ro=+\infty$, respectively. If one starts to move on the $u$-axis from the right to the left, from $u=+\infty$ ($r=0$) to $u_r^+$ to $u_r^-$ to $u_-$ to $\cdots$ to $u_2$  to $u_3$ and finally to $u=-\infty$. Since $\ro$ is an increasing function of $u$, the corresponding point on the $\ro$-axis starts to move from $\ro_0$ to $\ro=-\infty$ and then back in a circular rotation to $\ro=+\infty$ to $\ro_-$ to $\cdots$ to $e_2$ to $e_3$ and finally to $\ro_0$ again. Thus, we have the following order relations for the $\ro$-parameters:
\begin{equation}\label{A6}
   -\infty<\ro_0< e_3<\ri<e_2<e_1<\ro_+<\ro_-<\ro_r=+\infty\,.
\end{equation}

If $\ell=0$ ($M_2$ on the $u$-axis), then $u_r=u_-$, $u_2=u_3=0$ and $u_1=u_+$ ($\ro_r=+\infty$, $e_2=e_3$ and $e_1=\ro_+$). Using~(\ref{3.13}, \ref{5.5a}, \ref{5.5b}, \ref{5.6}, \ref{5.7}, \ref{5.10}, \ref{5.11}) we obtain $12\ro_0=5-[6\eta_2(1+\sqrt{1-\eta_2a^2})/a^2]$ (with $\eta_2=+1$), thus
\begin{equation}\label{A7}
  -\infty<\ro_0<e_3=e_2=\ri=-\frac{1}{12}<e_1=\ro_+=\frac{1}{6}<\ro_-=+\infty\,.
\end{equation}

If $\ell=\ell_+$, then $u_1=u_2$ ($M_3$ on the $u$-axis) and~\eqref{A6} becomes
\begin{equation}\label{A8}
   -\infty<\ro_0< e_3<\ri<e_2=e_1<\ro_+<\ro_-<\ro_r=+\infty\,.
\end{equation}

If $\ell=\ell_-<0$, there is still one real root $u_+<u_r=u_1<u_-$ ($M_1$ on the $u$-axis). This case is excluded, however, if it were possible for a photon to move with a negative energy, it could do it on a confined stable circle with radius $r_-<r=M(3-\sqrt{9-8a^2})/2<r_+$ which shrinks to zero as $a^2=q^2/M^2$ approaches zero.


\end{document}